\documentclass[aps,prd,onecolumn,groupedaddress,showpacs,nofootinbib,amssymb]{revtex4}
\usepackage[dvips]{graphicx}
\usepackage{amssymb}
\usepackage{amsmath}
\usepackage{graphicx,,color}
\usepackage{amsfonts}
\usepackage{bm}
\usepackage{cancel}
\usepackage{comment}

\newcommand\be{\begin{equation}}
\newcommand\ee{\end{equation}}

\allowdisplaybreaks[4]

\begin{document}

\tolerance=5000

\title{Rescaled Einstein-Gauss-Bonnet Gravity Inflation}
\author{V.K. Oikonomou$^{1,2}$}
\email{Corresponding author:
voikonomou@gapps.auth.gr;v.k.oikonomou1979@gmail.com}
\author{Ardit Gkioni$^{1}$}\,
\email{arisgkioni@gmail.com}
\author{Iason Sdranis$^{1}$}\,
\email{isdranis@protonmail.com}
\author{Pyotr Tsyba$^{2}$}\,
\email{pyotrtsyba@gmail.com}
\author{Olga Razina$^{2}$}\,
\email{olvikraz@mail.ru} \affiliation{$^{1)}$Department of
Physics, Aristotle University of Thessaloniki, Thessaloniki 54124,
Greece} \affiliation{$^{2)}$L.N. Gumilyov Eurasian National
University - Astana, 010008, Kazakhstan}

 \tolerance=5000

\begin{abstract}
We study the inflationary phenomenology of a rescaled
Einstein-Gauss-Bonnet gravity. In this framework, the
gravitational constant of the Einstein-Hilbert term is rescaled
due to effective terms active in the high curvature era.
Basically, the total theory is an $F(R,G,\phi)$ theory with the
Gauss-Bonnet part contributing only a non-minimal coupling to the
scalar field, so it is a theory with string theory origins and
with a non-trivial $F(R)$ gravity part. The $F(R)$ gravity part in
the high curvature regime contributes only a rescaled
Einstein-Hilbert term and thus the resulting theory is effectively
a rescaled version of a standard Einstein-Gauss-Bonnet theory. We
develop the formalism of rescaled Einstein-Gauss-Bonnet gravity,
taking in account the GW170817 constraints on the gravitational
wave speed. We show explicitly how the rescaled theory affects
directly the primordial scalar and tensor perturbations, and how
the slow-roll and observational indices of inflation are affected
by the rescaling of the theory. We perform a thorough
phenomenological analysis of several models of interest and we
show that is it possible to obtain viable inflationary theories
compatible with the latest Planck data. Also among the studied
models there are cases that yield a relatively large blue tilted
tensor spectral index and we demonstrate that these models can
lead to detectable primordial gravitational waves in the future
gravitational wave experiments. Some of the scenarios examined,
for specific values of the reheating temperature may be detectable
by SKA, LISA, BBO, DECIGO and the Einstein Telescope.
\end{abstract}

\pacs{04.50.Kd, 95.36.+x, 98.80.-k, 98.80.Cq,11.25.-w}

\maketitle

\section{Introduction}

One of the most refined mental perceptions of theoretical
physicists about the early Universe dynamics is the inflationary
paradigm \cite{inflation1,inflation2,inflation3,inflation4}, which
solves fundamental problems of the Big Bang cosmology in a refined
and elegant way. Also, its prospects of detection are realistic
and thus it is not a vague theoretical construction that yields
elegant physics. Indeed, the future Cosmic Microwave Background
(CMB) experiments, like the Simons observatory
\cite{SimonsObservatory:2019qwx} and hopefully the CMB stage 4
experiments \cite{CMB-S4:2016ple}, if these become operational,
will directly probe or constraint the $B$-modes of the CMB. These
$B$-modes constitute a direct probe for the primordial occurrence
of the inflationary era. Apart from that, the future gravitational
wave experiments
\cite{Hild:2010id,Baker:2019nia,Smith:2019wny,Crowder:2005nr,Smith:2016jqs,Seto:2001qf,Kawamura:2020pcg,Bull:2018lat,LISACosmologyWorkingGroup:2022jok}
can also provide concrete information for the existence of a
stochastic gravitational wave background of primordial origin.
These are tests for inflation, that will hopefully yield some
information about this mysterious primordial era of our Universe.
It is conceivable that the verification of the inflationary era
will be one of the most important achievements of theoretical
physics, and of equal importance with the discovery of the Higgs
particle, if not greater. The observational signs coming from
NANOGrav \cite{NANOGrav:2023gor} point out to the existence of a
stochastic gravitational wave background, which is not certain yet
whether it is of cosmological or astrophysical origin. However, in
the context of inflationary theories it is hard or even impossible
to explain the 2023 NANOGrav stochastic gravitational wave signal
\cite{Vagnozzi:2023lwo}, and inflation in conjunction with extra
physics might explain the signal see for example
\cite{Yi:2023mbm,Balaji:2023ehk,Oikonomou:2023qfz}. One is
certain, inflation alone cannot explain the NANOGrav signal.

With regard to the inflationary paradigm, there are two mainstream
descriptions, the single scalar field theory inflationary
description \cite{inflation1,inflation2,inflation3,inflation4},
and the geometric realization of inflation in the context of some
modified gravity \cite{reviews1,reviews2,reviews3,reviews4}. One
candidate theory that can describe efficiently the inflationary
era is Einstein-Gauss-Bonnet (EGB) theory of gravity
\cite{Hwang:2005hb,Nojiri:2006je,Cognola:2006sp,Nojiri:2005vv,Nojiri:2005jg,Satoh:2007gn,Bamba:2014zoa,Yi:2018gse,Guo:2009uk,Guo:2010jr,Jiang:2013gza,vandeBruck:2017voa,Pozdeeva:2020apf,Vernov:2021hxo,Pozdeeva:2021iwc,Fomin:2020hfh,DeLaurentis:2015fea,Chervon:2019sey,Nozari:2017rta,Odintsov:2018zhw,Kawai:1998ab,Yi:2018dhl,vandeBruck:2016xvt,Maeda:2011zn,Ai:2020peo,Easther:1996yd,Codello:2015mba},
which can actually predict a blue-tilted tensor spectral index, a
feature absent in single scalar field and $f(R)$ gravity
descriptions of inflation. The motivation for these theories is
strong, since the EGB theories are basically string theory
originating theories. Indeed, if one considers that a scalar field
is responsible for the inflationary dynamics, then the most
general two derivative scalar field Lagrangian in four dimensions
is the following,
\begin{equation}\label{generalscalarfieldaction}
\mathcal{S}_{\varphi}=\int
\mathrm{d}^4x\sqrt{-g}\left(\frac{1}{2}Z(\varphi)g^{\mu
\nu}\partial_{\mu}\varphi
\partial_{\nu}\varphi+\mathcal{V}(\varphi)+h(\varphi)\mathcal{R}
\right)\, .
\end{equation}
The evaluation of the scalar field in its vacuum configuration,
imposes the constraints that the scalar field can be either
conformally coupled or minimally coupled to gravity. The quantum
corrections to the above gravitational action which are also
consistent with diffeomorphism invariance is
\cite{Codello:2015mba},
\begin{align}\label{quantumaction}
&\mathcal{S}_{eff}=\int
\mathrm{d}^4x\sqrt{-g}\Big{(}\Lambda_1+\Lambda_2
\mathcal{R}+\Lambda_3\mathcal{R}^2+\Lambda_4 \mathcal{R}_{\mu
\nu}\mathcal{R}^{\mu \nu}+\Lambda_5 \mathcal{R}_{\mu \nu \alpha
\beta}\mathcal{R}^{\mu \nu \alpha \beta}+\Lambda_6 \square
\mathcal{R}\\ \notag &
+\Lambda_7\mathcal{R}\square\mathcal{R}+\Lambda_8 \mathcal{R}_{\mu
\nu}\square \mathcal{R}^{\mu
\nu}+\Lambda_9\mathcal{R}^3+\mathcal{O}(\partial^8)+...\Big{)}\, ,
\end{align}
where the parameters $\Lambda_i$, $i=1,2,...,6$ are appropriate
dimensionful constants. Apparently the EGB theories of gravity are
a subclass of these quantum corrections. The attributes of EGB
theories have been frequently studied and presented in the
relevant literature
\cite{Hwang:2005hb,Nojiri:2006je,Cognola:2006sp,Nojiri:2005vv,Nojiri:2005jg,Satoh:2007gn,Bamba:2014zoa,Yi:2018gse,Guo:2009uk,Guo:2010jr,Jiang:2013gza,vandeBruck:2017voa,Pozdeeva:2020apf,Vernov:2021hxo,Pozdeeva:2021iwc,Fomin:2020hfh,DeLaurentis:2015fea,Chervon:2019sey,Nozari:2017rta,Odintsov:2018zhw,Kawai:1998ab,Yi:2018dhl,vandeBruck:2016xvt,Maeda:2011zn,Ai:2020peo,Easther:1996yd,Codello:2015mba}.
However in 2017 the striking GW170817 event
\cite{TheLIGOScientific:2017qsa,Monitor:2017mdv,GBM:2017lvd,LIGOScientific:2019vic}
has cast doubts on theories predicting  propagation speed for
tensor perturbations distinct from that of the light speed
\cite{Ezquiaga:2017ekz,Baker:2017hug,Creminelli:2017sry,Sakstein:2017xjx,Boran:2017rdn}.
The remedy for EGB theories was given in a series of articles
\cite{Oikonomou:2021kql,Oikonomou:2022xoq,Odintsov:2020sqy}, in
which it was shown that the resulting EGB theories can be
compatible with the GW170817 event, if the scalar potential the
EGB scalar coupling function are constrained and related in a
specific way. In this work, we aim to study these viable classes
of EGB theories, with the fundamental difference that gravity is
stronger or weaker during the inflationary era. We call these
theories rescaled EGB theories, since the Einstein-Hilbert term is
rescaled by a constant number. Such rescaled EGB theories are
effective theories originating possible by higher order terms of
the Ricci scalar in the action, which during the early time era
yield such a rescaling in the Einstein-Hilbert term. These
theories are motivated from fundamental principles, since the
effective Lagrangian of $f(R)$ gravity is in general dependent on
basic cosmological parameters like the cosmological constant and
the same $f(R)$ gravity has to describe the early and late-time
eras of our Universe. So if $f(R)$ gravity is the optimal theory
for describing inflation and dark energy, then effective
exponential terms of the form $\lambda e^{-\Lambda/R}$ are
expected to appear in the theory, where $\Lambda$ is the
cosmological constant. Such terms in the large curvature limit
lead to terms of the form $\lambda R$, so effectively, the
primordial $f(R)$ gravity during inflation is a rescaled
Einstein-Hilbert gravity. Now a good question is whether
instabilities or ghost degrees of freedom arise in the theory.
However, no such issue arises since the propagating degrees of
freedom of the effective rescaled theory is the same as in
ordinary $f(R)$ gravity in the presence of EGB terms. The only
problematic issue is the possibility to have antigravity effects
if the rescaling $\lambda R$ is negative and larger than $R$, but
this can avoided by suitably choosing the parameter $\lambda$.We
will examine in detail the inflationary phenomenology of such
theories, pointing out the significance of the rescaling and also
we point out the cases in which the tensor spectral index acquires
large values, a feature important for future gravitational wave
detections. We examine several models of theoretical importance
and we confront the results with the latest Planck data.
Furthermore, for the models that predict a blue-tilted tensor
spectrum, we evaluate the predicted energy spectrum of the
primordial gravitational waves and we confront the findings with
the sensitivity curves of the future and current gravitational
waves experiments. As we show, the predicted energy spectrum of
primordial gravitational waves from rescaled EGB gravities can be
detectable in future gravitational wave experiments.

\section{The Rescaled EGB Theoretical Framework}

The rescaled EGB theory of gravity has the following gravitational
action,
\begin{equation}
\centering \label{action} \mathcal{S}=\int
d^4x\sqrt{-g}\bigg(\frac{\beta
R}{2\kappa^2}-\frac{1}{2}g^{\mu\nu}\nabla_\mu\phi\nabla_\nu\phi-V(\phi)-\xi(\phi)\mathcal{G}\bigg)\,
,
\end{equation}
with $\kappa=\frac{1}{M_{Pl}}=\sqrt{8\pi G}$ being the
gravitational constant, $M_{Pl}$ standing for the reduced Planck
mass and $\beta$ is a dimensionless constant, which quantifies the
rescaling of the Einstein-Hilbert term. Such rescaled versions of
EGB gravity may originate from higher order $f(R)$ gravities, with
action,
\begin{equation}
\label{action1231} \centering \mathcal{S}=\int d^4x\sqrt{-g}
\frac{f(R)}{2\kappa^2}\, .
\end{equation}
One example of such effective $f(R)$ gravities is
\cite{Oikonomou:2020oex},
\begin{equation}\label{frini}
f(R)=R-\gamma  \lambda  \Lambda -\lambda  R \exp
\left(-\frac{\gamma  \Lambda }{R}\right)-\frac{\Lambda
\left(\frac{R}{m_s^2}\right)^{\delta }}{\zeta }\, ,
\end{equation}
which in the large curvature limit yields the following form,
\begin{equation}\label{expapprox}
\lambda  R \exp \left(-\frac{\gamma  \Lambda }{R}\right)\simeq
-\gamma \lambda  \Lambda -\frac{\gamma ^3 \lambda \Lambda^3}{6
R^2}+\frac{\gamma ^2 \lambda  \Lambda ^2}{2 R}+\lambda  R\, ,
\end{equation}
therefore, the complete effective action when the curvature is
high takes the form,
\begin{equation}\label{effectiveaction}
\mathcal{S}=\int d^4x\sqrt{-g}\left(\frac{1}{2\kappa^2}\left(\beta
R+ \frac{\gamma ^3 \lambda \Lambda ^3}{6 R^2}-\frac{\gamma ^2
\lambda \Lambda ^2}{2 R}-\frac{\Lambda}{\zeta
}\left(\frac{R}{m_s^2}\right)^{\delta
}+\mathcal{O}(1/R^3)+...\right)-\frac{1}{2}g^{\mu\nu}\nabla_\mu\phi\nabla_\nu\phi-V(\phi)-\xi(\phi)\mathcal{G}\right)\,
,
\end{equation}
where $\beta=1-\lambda$. Note that the effective Newton's constant
for these theories is distinct from the ordinary value $G$, so
this could potentially affect the Big Bang Nucleosynthesis (BBN).
However, this line of thinking is not correct, because the
effective rescaled theory is valid only when the curvature is
high. Thus post-inflationary and even during the reheating era,
the rescaled action is no longer valid and the original action
should be valid. Thus taking the dominant leading order terms in
the action (\ref{effectiveaction}), we will start our analysis
with the following action,
$$
S=\int{}^{}d^4x\left(\beta{}\frac{R}{2\kappa^2} +
\frac{1}{2}(\partial_\mu \phi)(\partial^\mu \phi) -V(\phi)
-\frac{1}{2}\xi(\phi)\mathcal{G}\right) \hspace{0.5cm}, $$
with
$$\mathcal{G}= R^2
-4R_{\alpha\beta}R^{\alpha\beta}+R_{\alpha\beta\gamma\delta}R^{\alpha\beta\gamma\delta}\,
.$$

Furthermore, it is assumed that the background metric is described
by the Friedmann-Robertson-Walker (FRW) metric which is:
$$ds^2 = -dt^2 + a^2(t)\sum_{i=1}^{3}(dx^i)^2 \hspace{0.4cm}, $$
where in this metric framework , $a(t)$ represents the scale
factor, and the Gauss-Bonnet invariant takes the form
$\mathcal{G}=24H^2 (\dot{H} + H^2)$. Upon varying the action with
respect to the metric, for a general metric we obtain the
following field equations for the rescaled EGB gravity.
\begin{equation}
\centering \label{fieldeq} \frac{\beta}{\kappa^2}
G_{\mu\nu}=T^{(\xi)}_{\mu\nu}+\nabla_\mu\phi\nabla_\nu\phi-\bigg(\frac{1}{2}g^{\alpha\beta}\nabla_\alpha\phi\nabla_\beta\phi+V\bigg)g_{\mu\nu}\,
,
\end{equation}
with the stress-energy tensor corresponding to the EGB string
corrections being
$T_{\mu\nu}^{(\xi)}=-\frac{2}{\sqrt{-g}}\frac{\delta(\sqrt{-g}\xi(\phi)\mathcal{G})}{\delta
g^{\mu\nu}}$ and it is equal to,
\begin{widetext}
\begin{align}
\centering \label{Txi} &
T^{(\xi)}_{\mu\nu}=-2\bigg[\bigg(\frac{1}{2}\mathcal{G}g_{\mu\nu}+4R_{\mu\alpha}R^\alpha_\nu+4R^{\alpha\beta}R_{\mu\alpha\nu\beta}-2R_{\mu}^{\,\,\,\alpha\beta\gamma}R_{\nu\alpha\beta\gamma}-2RR_{\mu\nu}\bigg)\xi
\\ \notag &-4\bigg(\xi^{;\alpha\beta}R_{\mu\alpha\nu\beta}-\Box\xi
R_{\mu\nu}+2\xi_{;\alpha(\nu}R^\alpha_{\mu)}-\frac{1}{2}\xi_{,\mu;\nu}R\bigg)+2(2\xi_{;\alpha\beta}R^{\alpha\beta}-\Box\xi
R)g_{\mu\nu}\bigg]\, .
\end{align}
\end{widetext}
Accordingly, variation of the action with respect to the scalar
field yields,
\begin{equation}
\centering \label{conteq} \Box\phi-V'-T^{(\xi)}=0\, ,
\end{equation}
where $\Box=\nabla_\mu\nabla^\mu$ and $T^{(\xi)}=\xi'\mathcal{G}$.
For the FRW metric, the field equations read,
\begin{gather}
\frac{3\beta H^2}{\kappa^2} = \frac{1}{2}\dot{\phi}^2 + V + 12\dot{\xi}H^3 \label{eq:1} \\ \notag \\
\frac{2\beta\dot{H}}{\kappa^2} = -\dot{\phi}^2 + 4\ddot{\xi}H\dot{H} - 4\dot{\xi}H^3 \label{eq:2} \\ \notag \\
\ddot{\phi}+3H\dot{\phi}+V' + 12\xi'H^2(\dot{H}+H^2) =0 \label{eq:3}
\end{gather}
At this point it is apparent how the rescaled Einstein-Hilbert
term $\beta R$ affects the field equations, since the rescaling
parameter $\beta$ already appears in the field equations
(\ref{eq:1}) and (\ref{eq:2}). As we will show shortly, the
rescaling parameter $\beta$ affects the slow-roll indices too.
Considering the inflationary era of the EGB theory , we shall work
under the following slow-roll conditions:
\begin{gather}
\dot{H}\ll{}H^2 \hspace{0.2cm},\hspace{0.2cm} \frac{\dot{\phi}^2}{2}\ll{}V \hspace{0.1cm},\hspace{0.2cm} \ddot{\phi}\ll{}3H\dot{\phi} \label{eq:4}
\end{gather}
The slow-roll conditions have no direct relation with the
rescaling of the Einstein-Hilbert term. These slow-roll conditions
are basically the requirement that a slow-roll era of inflation
occurs and these relations are not affected at all by the
rescaling. Hence the slow-roll conditions are identical for the
ordinary EGB and the rescaled EGB theory. Calculations will be
performed under the premise, that the gravitational wave of tensor
perturbations \cite{Hwang:2005hb},
\begin{gather}c^2_T=1-\frac{Q_f}{2Q_t} \label{eq:5} \end{gather}
is equal to unity in natural units. The functions above are,
$Q_f=8(\ddot{\xi}-H\dot{\xi}),\hspace{0.1cm}
Q_t=F+\frac{Q_b}{2},\hspace{0.1cm}F=\frac{\beta}{\kappa^2}
\hspace{0.1cm}and\hspace{0.1cm} Q_b=-8\dot{\xi}H.$ The condition
$c^2_T=1$ demands that $Q_f=0$ which results in a differential
equation constraining the Gauss-Bonnet scalar coupling function
$\xi(\phi)$. In terms of the scalar field the above differential
equation, given that $\dot{\xi}=\xi'\dot{\phi}$, is written as
follows,
\begin{gather}
\xi''\dot{\phi}^2+\xi'\ddot{\phi}=H\xi'\dot{\phi}\label{eq:6}
\end{gather}
Motivated by the scalar field's slow-roll conditions, we further assume,
\begin{gather}
\dot{\phi}\simeq{}\frac{H\xi'}{\xi''} \label{eq:7}
\end{gather}
Combination of the above equations yields,
\begin{gather}
\frac{\xi'}{\xi''}\simeq{}=-\frac{1}{3H^2}(V'+12\xi'H^4)\label{eq:8}
\end{gather}
We are now able to express all the slow roll indices with respect
to the $\frac{\xi'}{\xi''}$ ratio, while upon combining,
\begin{gather}
\kappa\frac{\xi'}{\xi''}\ll{}1 \label{eq:9}\\12\dot{\xi}H^3=12\frac{\xi'^2H^4}{\xi''}\ll{}V \label{eq:10}
\end{gather}
we can further simplify the equations of motion:
\begin{gather}
\beta{}H^2\simeq{}\frac{\kappa^2V}{3} \label{eq:11}\\ \vspace{0.4cm}
\beta{}\dot{H}\simeq{}-\frac{1}{2}\kappa^2\dot{\phi}^2 \label{eq:12}\\ \vspace{0.4cm}
\dot{\phi}\simeq{}\frac{H\xi'}{\xi''} \label{eq:13}
\end{gather}
Having generated a simplified version of the field equations, we
get,
\begin{gather}
\frac{4\kappa^2\xi'^2V}{3\beta{}^2\xi''}\ll{}1 \label{eq:14}
\end{gather}an approximation, which will be employed in the simplification of the slow-roll indices' expressions, starting with our initial differential equation of the scalar coupling function $\xi(\phi)$ , which becomes,
\begin{gather}
\beta{}^2\frac{V'}{V^2}+\frac{4\kappa^4}{3}\xi'\simeq{}0 \label{eq:15}
\end{gather}
Equation (\ref{eq:15}) connects the scalar coupling function
$\xi(\phi)$ with the inflationary potential $V(\phi)$, which will
be of paramount importance for the analysis that follows.
Recalling the definitions of the slow-roll indices
\cite{Hwang:2005hb},
\begin{gather}
\epsilon_1=-\frac{\dot{H}}{H^2} ,\hspace{0.2cm} \epsilon_2=\frac{\ddot{\phi}}{H\dot{\phi}},\hspace{0.2cm} \epsilon_3=\frac{\dot{F}}{2HF},\hspace{0.2cm}\epsilon_4=\frac{\dot{E}}{2HE},\hspace{0.2cm} \epsilon_5=\frac{\dot{F}+Q_a}{2HQ_t}, \hspace{0.2cm} \epsilon_6=\frac{\dot{Q_t}}{2HQ_t}, \label{eq:16}
\end{gather}
with $F=\frac{\partial{}F}{\partial{}R}=\frac{\beta{}}{\kappa^2}$ and \textit{E} defined as ,
\begin{gather}
E=\frac{F}{\dot{\phi}^{2}}\left(\dot{\phi}^2+3\left(\frac{\left(\dot{F}+Q_a^2\right)}{2Q_t}\right)\right) \label{eq:17}
\end{gather}
whereas the $Q_i$ functions are given by \cite{Hwang:2005hb},
\begin{gather}
Q_a=-4\dot{\xi}H^2,\hspace{0.2cm}Q_b=-8\dot{\xi}H,\hspace{0.2cm}Q_t=F+\frac{Q_b}{2},\hspace{0.2cm} Q_e=16\dot{\xi}\dot{H} \label{eq:18} \\ \notag \\
Q_a=-4\frac{\xi^{'2}}{\xi^{''}}\left(\frac{\kappa^{2}V}{3\beta{}}\right)^{3/2},\hspace{0.2cm}Q_b=-\frac{8\kappa^{2}\xi^{'2}V}{3\xi^{''}\beta{}},\hspace{0.2cm}Q_t=\frac{\beta{}}{\kappa^{2}}\left(1-\frac{4\kappa^{4}\xi^{'2}V}{3\xi^{''}\beta^{2}}\right),\hspace{0.2cm} Q_e=16\frac{\xi^{'3}}{\xi^{''2}}H^{2}H^{'}
\end{gather}
Applying the equations of motion relating $H$, $\dot{H}$ to the
$\frac{\xi'}{\xi''}, V(\phi)$ we obtain:
\begin{align}
\epsilon_1&\simeq{}\frac{\kappa^2}{2\beta{}}\left(\frac{\xi'}{\xi''}\right)^2 \label{eq:19} \\ \notag \\
\epsilon_2&\simeq{}1-\epsilon_1-\frac{\xi'\xi'''}{\xi''^2} \label{eq:20}\\ \notag \\
\epsilon_3&=0 \label{eq:21} \\ \notag \\
\epsilon_4&\simeq{}\frac{\xi'}{2\xi''}\frac{E'}{E} \label{eq:22} \\ \notag \\
\epsilon_5&\simeq{}-\frac{\epsilon_{1}}{\lambda{}} \label{eq:23} \\ \notag  \\
\epsilon_6&\simeq{}\epsilon_{5}\left(1-\epsilon_{1}\right)\label{eq:24}
\end{align}
whereby the explicit forms of the $E({\phi})$ and assisting
$\lambda(\phi)$ function are,
\begin{gather}
E(\phi)=\frac{\beta{}}{\kappa^2}+\frac{8\kappa^4\xi'^2V^2}{3\beta{}^2\left(1-\frac{4\kappa^{4}\xi^{'2}V}{3\beta^{2}\xi^{''}}\right)} \label{eq:25} \\ \notag \\
\lambda(\phi)=\frac{3\beta{}}{4\kappa^{2}\xi^{''}V}
\end{gather}
 It is apparent at this point that the rescaling
parameter $\beta$ affects directly the slow-roll parameters, since
it appears explicitly in the first slow-roll index in Eq.
(\ref{eq:19}) and implicitly in the rest of the slow-roll indices,
via  $E({\phi})$ and $\lambda{}$ which contain the rescaling
parameter $\beta$.

The above definitions of the slow-roll indices allow for a
simplified representation of the observable quantities, namely the
spectral indices of the scalar and tensor primordial
perturbations, which are \cite{Hwang:2005hb},
\begin{align}
n_{\mathcal{S}}&=1-4\epsilon_1-2\epsilon_2-2\epsilon_4
\label{eq:26} \\ \vspace{0.8cm} n_{\mathcal{T}}&=-2(\epsilon_1 +
\epsilon_6)\label{eq:27}
\end{align}
while the tensor-to-scalar ratio is equal to,
\begin{gather}
\vspace{0.5cm} r=16\left|\left(\frac{\kappa^2Q_e}{4H}-\epsilon_1\right)\frac{2c^3_A}{2+\kappa^2Q_b}\right| \label{eq:28} \\ \vspace{0.5cm} c^2_A=1+\frac{Q_aQ_e}{3Q^2_a+\dot{\phi}^2\left(\frac{2}{\kappa^2}+Q_b\right)} \label{eq:29}
\end{gather}
with $c_A$ denoting the speed of sound for the scalar
perturbations. The detailed derivation of the final forms of the
observational indices and of the sound speed given in Eqs.
(\ref{eq:27})-(\ref{eq:29}) can be found in Ref.
\cite{Hwang:2005hb}. Lastly, this framework allows us to express
the number of \textit{e}-foldings in terms of the scalar field.
Given $\phi_i$ and $\phi_f$ the field values at the first horizon
crossing and at the end of inflation respectively, the number of
\textit{e}-foldings is,
\begin{gather}
N=\int_{t_i}^{t_f}Hdt=\int_{\phi_i}^{\phi_f}\frac{H}{\dot{\phi}}dt=\int_{\phi_i}^{\phi_f}\frac{\xi''}{\xi'}d\phi \label{eq:30} \vspace{0.3cm}
\end{gather}

\section{EBG Inflationary Phenomenology}

In this section, we delve further into detailed calculations of
slow-roll and scalar/tensor perturbation spectral indices for
candidate potentials $V(\phi)$ that can realize an inflationary
era. The models will be compared to the Planck data
\cite{Planck:2018jri}, which indicate that,
\begin{gather}
n_{\mathcal{S}}=0.9649\pm{}0.0042 , \hspace{0.3cm} r<0.064 \label{eq:31}
\end{gather}
Also we shall compare the results with the BICEP/Keck updates on
the Planck data \cite{BICEP:2021xfz} which constrain the
tensor-to-scalar ratio to be $r<0.036$ at $95\%$ confidence. For
further compatibility with the aforementioned data, values of the
tensor spectral index which are positive $n_{\mathcal{T}}>0$ will
be considered prime candidates to support the inflationary regime
described by our theory. In total, six candidate potentials
$V(\phi)$ will be tested out, in hopes of reproducing blue-tilted
values of the spectral and index, as well as the tensor-to-scalar
ratio value. Furthermore, we would like to achieve these results
with simultaneous compatibility with the 2018 Planck data.

\subsection{Model I: Power law choice for $V(\phi)$}

\begin{figure}
    \centering
    \includegraphics[width=0.45\textwidth]{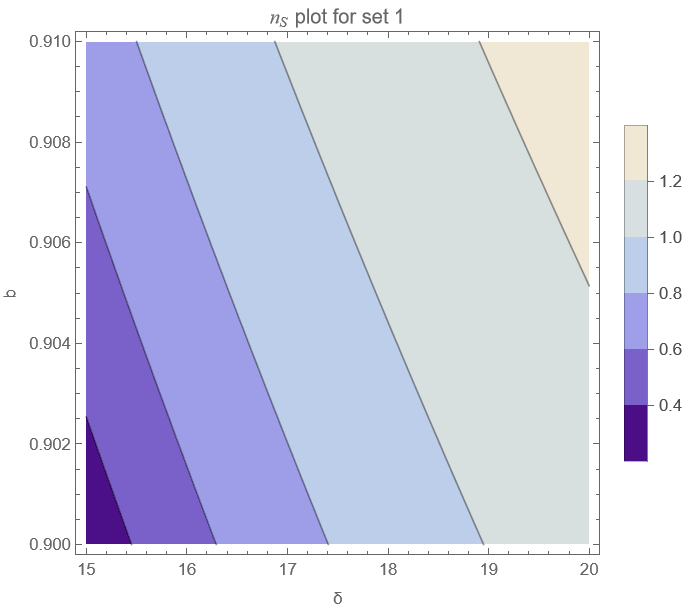}
    \vspace{0.5em}
    \includegraphics[width=0.45\textwidth]{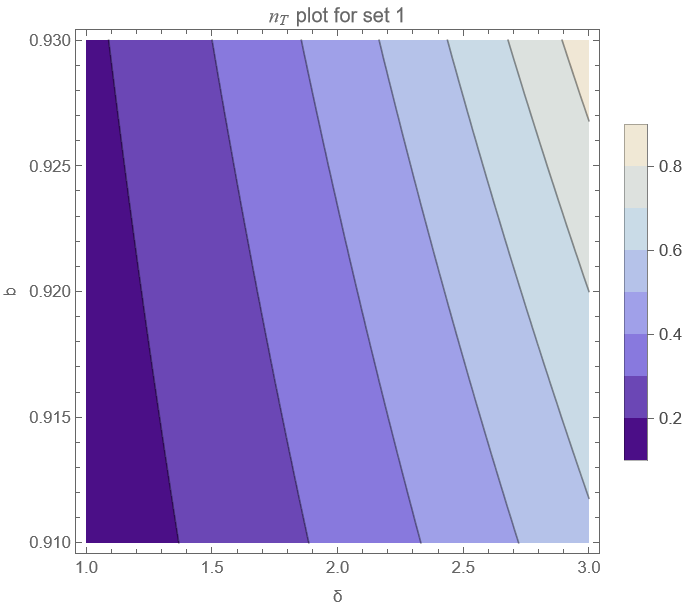}
\caption{The spectral indices for the free parameters of data Set
1 of our theory. Note that in each scenario, the indices respect
the 2018 Planck constraints on specific contour curves of
$n_{\mathcal{S}}$, yet in each scenario they yield a blue-tilted
$n_{\mathcal{T}}$ and respect the constraints we've imposed . The
same analysis will follow for the rest of the data Sets, but the
same behavior should be the case for the rest of the data Sets.}
\label{fig:3}
\end{figure}

\begin{figure}
    \centering
    \includegraphics[width=0.45\textwidth]{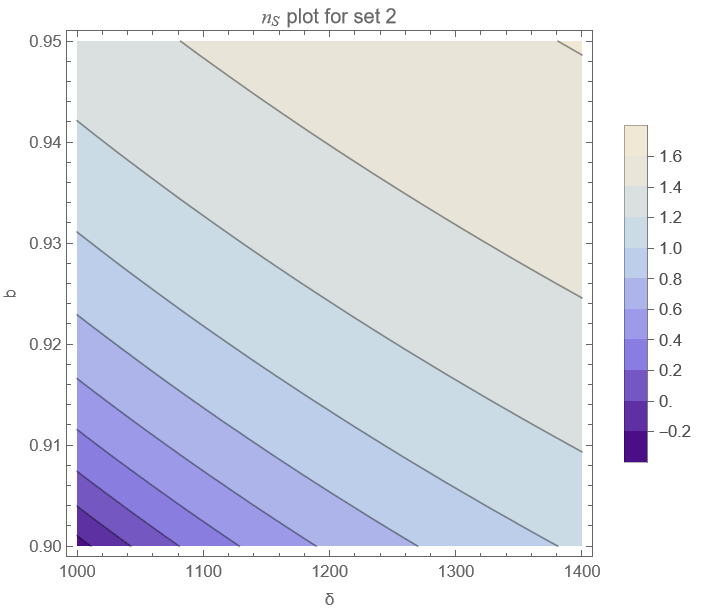}
    \includegraphics[width=0.45\textwidth]{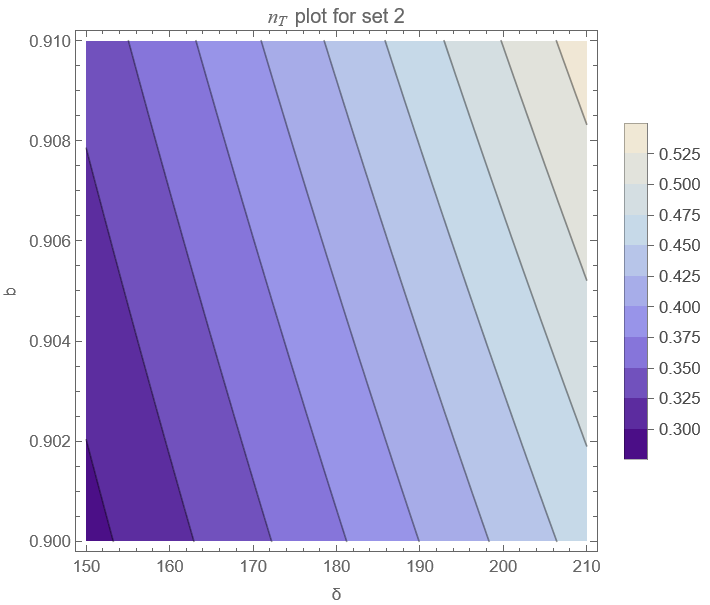}
\caption{The spectral indices for the free parameters of data Set
2 of our theory. Note the similar variations of both spectral
indices with the previous set and the blue-tilt that they commonly
result in.} \label{fig:4}
\end{figure}

\begin{figure}
    \centering
    \includegraphics[width=0.45\textwidth]{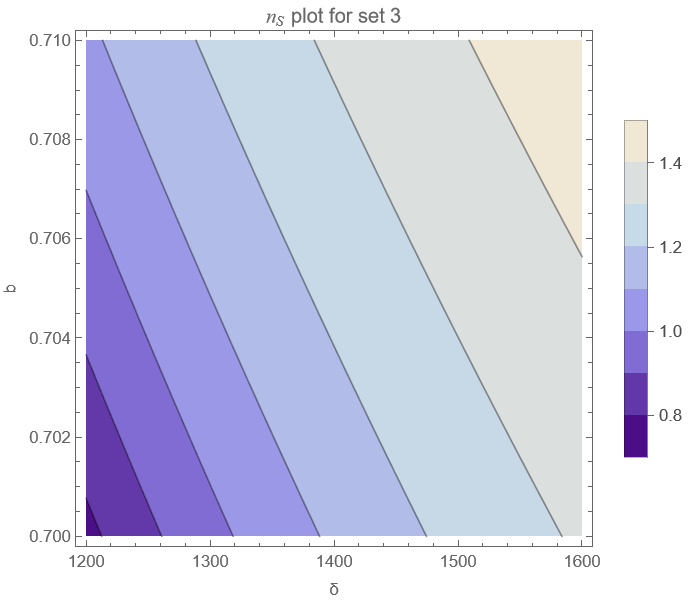}
    \includegraphics[width=0.45\textwidth]{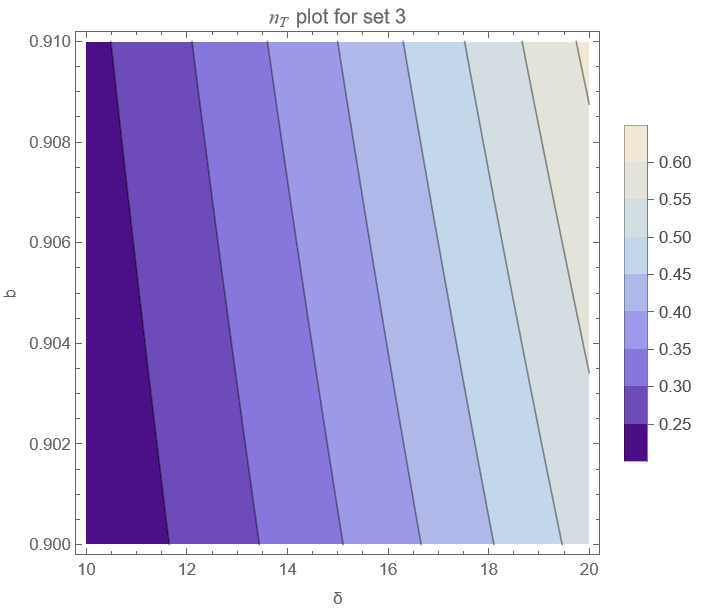}
\caption{The spectral indices for the free parameters of data Set
3 of our theory. It is immediately noted how the free parameters
alter the domain of satisfaction of the desired constraints.}
 \label{fig:7}
\end{figure}

\begin{figure}
    \centering
    \includegraphics[width=0.45\textwidth]{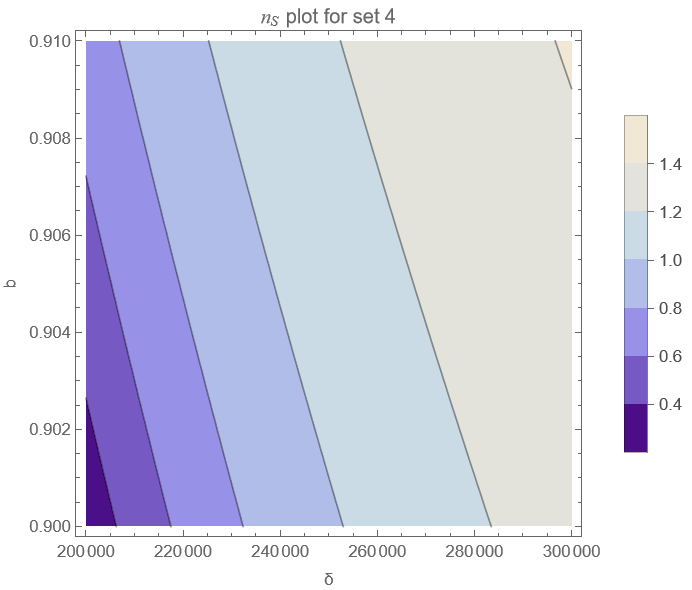}
    \includegraphics[width=0.45\textwidth]{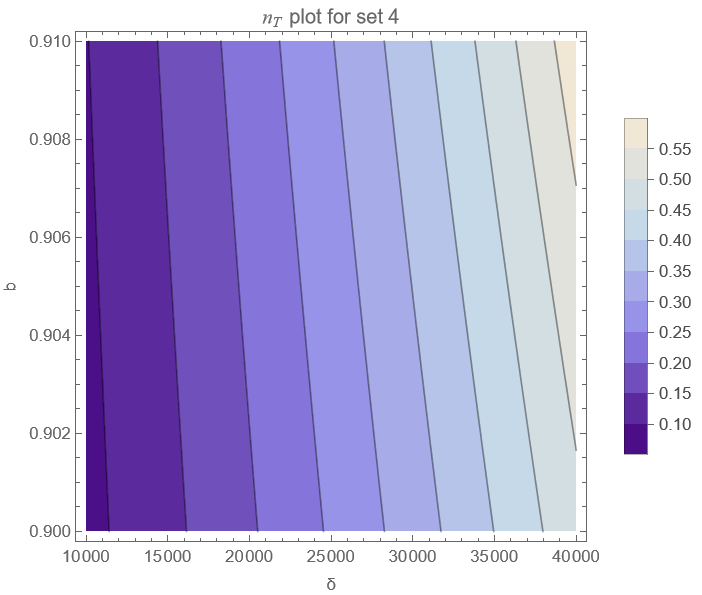}
\caption{The spectral indices for the free parameters of data Set
4 of our theory.} \label{fig:9}
\end{figure}

\begin{figure}[h!]
\centering
\includegraphics[width=20pc]{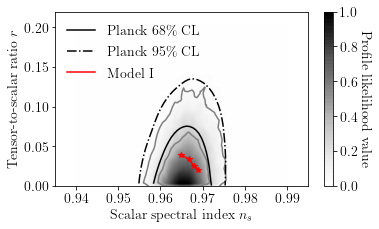}
\caption{The phenomenology of model I confronted with the Planck
2018 likelihood curves for various optimal values of the free
parameters.} \label{likelihoodmodel1}
\end{figure}

In this scenario , the simplest example of an inflationary
potential is considered,
\begin{gather}
 \xi(\phi)=\delta{}(\kappa\phi)^\nu{} \label{eq:32}
\end{gather}
where $\delta$ is a  parameter of dimension
\begin{gather}
\left[\delta{}\right]=\left[m\right]^{0}
\end{gather}
as it follows from $\left[R\right]=\left[m\right]^{-2}$ thus
$\mathcal{G}\propto{}R^{2}\Rightarrow{}\left[\mathcal{G}\right]=\left[m\right]^{-4}$
and $\kappa=\frac{1}{M_p}$ being the reciprocal of the Planck
mass. The corresponding potential follows from Eq. (\ref{eq:15}),
\begin{gather}
V(\phi)=\frac{3\beta{}^2}{4\delta \kappa^{4} (\kappa \phi)^\nu
+4\kappa^4\gamma} \label{eq:33}
\end{gather}
where $\gamma$ is a dimensionless integration constant. It is
noteworthy, that the addition of a factor $\gamma{}$ to the
potential has been implemented for the solemn purpose of avoiding
irregular behavior around $\phi=0$. As such, a phase tilt will
bear no significant effect to the collective phenomenon. At this
point, the slow roll indices are evaluated,
\begin{align}
\epsilon_{1}&=\frac{\kappa^{2}\phi^{2}}{2\beta{}\left(\nu{}-1\right)^{2}} \\ \notag \\
\epsilon_{2}&=-\frac{\kappa^{2} \phi^{2}}{2\beta{}(\nu{}-1)^{2}}-\frac{\nu{}-2}{\nu{}-1}+1 \\ \notag \\
\epsilon_{3}&=0 \\ \notag
\end{align}
\begin{align}
\epsilon_{4}&=\frac{\xi^{'}}{2\xi^{''}}\frac{\mathcal{E}^{'}}{\mathcal{E}} \\ \notag \\
\epsilon_{5}&\simeq{}-\frac{2\kappa^{2}\xi^{'2}V}{3\beta^{2}\xi^{''}} \\ \notag \\
\epsilon_{6}&\simeq{}\frac{2\kappa^{2}\xi^{'2}V}{3\beta^{2}\xi^{''}}\left(1-\frac{\kappa^{2}}{2\beta{}}\left(\frac{\xi^{'}}{\xi^{''}}\right)^{2}\right)
\end{align}
$n_{\mathcal{S}}$, $n_{\mathcal{T}}$ and $r$ which are of
importance are equally as complicated and rather inelegant in
their expressions, hence why they have been omitted from being
presented. In order to evaluate the above indices at the first
horizon crossing, recalling that $\phi_f$ and $\phi_i$ are
connected via the number of $e$-foldings relation (\ref{eq:30}) we
start by finding the scalar field value at the end of the
inflationary period, characterized by $\epsilon_1=1$. Without the
necessity of approximations, we obtain
$\phi_f=\frac{\sqrt{2\beta{}}(\nu{}-1)}{\kappa}$ and thus the
initial scalar field value being
$\phi_i=\frac{\sqrt{2\beta{}}(\nu{}-1)e^{-\frac{N}{\nu{}-1}}}{\kappa{}}$.\hspace{0.2cm}At
this value, the spectral indices pertaining to our framework are
expressed as complicated relations for which we have evaluated
using code. To better grasp the $n_{\mathcal{S}}$,
$n_{\mathcal{T}}$ and $r$ dependency on the dimensionless
parameters $\delta{}, \gamma{}, \nu{}$ as well as the new Ricci
scalar coupling $\beta{}$, we considered four sets of values,
listed in (\ref{set1}).
\begin{align}\label{set1}
set \hspace{0.2cm} 1 &: \nu{}=21 , \hspace{0.2cm} \gamma{}=5768 \vspace{0.8cm} \\
set \hspace{0.2cm} 2 &: \nu{}=19 , \hspace{0.2cm} \gamma{}=57  \vspace{0.8cm} \\
set \hspace{0.2cm} 3 &: \nu{}=19 , \hspace{0.2cm} \gamma{}=5  \vspace{0.8cm} \\
set \hspace{0.2cm} 4 &: \nu{}=21 , \hspace{0.2cm} \gamma{}=7.7\times{}10^8
\end{align}
Let us proceed by examining the dependence of
\textbf{$n_{\mathcal{S}}(\phi_i)$} in terms of
$(\beta{},\delta{})$ , the Ricci coupling and the power-law model
coefficient, in the parameter set 1 appearing in Eq. (\ref{set1}).
Assuming the number of \textit{e}-folds to be $N\simeq{}60$,
evaluating the spectral indices of primordial perturbations at the
first horizon crossing, for each set of parameters at a time,
yields the $n_{\mathcal{S}}$ contour plots appearing in Fig.
\ref{fig:3}.
So far, the graphs point towards optimistic results in regards to
a blue-tilted, rescaled Einstein-Gauss-Bonnet gravity for which
$\beta{}=1$ would reduce to the usual EGB framework. Another
noteworthy observation consists of the influence of the
dimensionless $\gamma{}$ parameter that was added as an
integration constant on the range of the potential's coefficient
$\delta{}$ for which the 2018 Planck constraints are satisfied. A
direct behavior cannot be deduced immediately, as both
$\left(\gamma{},\nu{}\right)$ parameters exert their influence on
the variation but we may view the respective qualitative and
quantitative behavior in Figs. \ref{fig:4}, \ref{fig:7} and
\ref{fig:9}.
At this point, we should be concerning ourselves with the prospect
of such a model potential yielding a blue-tilt in its tensor
spectral index of primordial perturbations with simultaneous
compatibility with the 2018 Planck constraints. In  Table
\ref{tab:1}, we've handpicked some arbitrary values of
$\left(\delta{},\beta{}\right)$ that respect the aforementioned
constraints and have been by the contours of Figs. \ref{fig:4},
\ref{fig:7} and \ref{fig:9}. The results are listed in Table
\ref{tab:1}. As it can be seen in Table \ref{tab:1}, the model is
compatible with the Planck data but also with the recent
BICEP/Keck data  \cite{BICEP:2021xfz}, and only the set 1 is not
compatible with the BICEP/Keck data.
\begin{table}[htbp]
    \centering
    \begin{tabular}{|c|c|c|c|c|c|} 
        \hline
        &   $\delta{}$&$\beta{}$&$n_{\mathcal{S}}$& $n_{\mathcal{T}}$& $r$\\ 
        \hline
        Set 1&   18&0.9&0.966& 0.08& 0.0396\\ \hline  
        Set 2&   4.63$\cdot10^{7}$&0.25&0.966& 0.56& 0.020\\ \hline  
        Set 3&   25&0.6&0.966& 0.035& 0.020\\ \hline  
         Set 4&   2.48$\cdot{}10^{5}$&0.725&0.966& 0.428& 0.0396\\\hline
    \end{tabular}
    \caption{Validity of a blue-tilted regime for Power Law Inflation, accounting for different data Sets and free-parameter values}
    \label{tab:1}
\end{table}
Furthermore, as it can be seen in Table \ref{tab:1}, the set 2 and
set 4 produce a significant blue tilted tensor spectrum while
simultaneously being compatible with the Planck and the BICEP/Keck
constraints. In addition, in Fig. \ref{likelihoodmodel1} we
confront model I with the Planck 2018 likelihood curves. As it can
be seen, the model is nicely fitted in the Planck likelihood
curves. We need to note here that the Planck data do not directly
constrain the tensor spectral index because there is no detection
of tensor modes at all. There are the LIGO-Virgo constraints on
the tensor spectral index, which are taken into account in a later
section, but still these are too restrictive to constrain the
tensor spectral index for  large frequency range starting from the
LIGO-Virgo operational frequencies down to the Litebird
frequencies. The reason is that a broken power-law behavior is
expected to apply in these frequencies, as discussed in Ref.
\cite{Benetti:2021uea}. Also it is noticeable that the rescaling
parameter $\beta$ plays an important role since it allows a great
diversity in the choice of free parameters and a diversity in the
viable inflationary phenomenologies. The only remaining feature to
attest to, is the validation of the slow-roll approximation regime
which we have assumed to hold true throughout our entire analysis.
In essence, this requires that for all of the above parameter sets
and $\left(\delta{},\beta{}\right)$ values, the conditions imposed
slow-roll conditions of the previous section are fully respected.
Starting with the very first set of parameters ,for the sake of
providing the process of validation, the condition (\ref{eq:4})
which is equivalent to requiring $\epsilon_{1}\ll{}1$ holds true
as for the 1st set of values it is of the order
$\mathcal{O}(10^{-7})$,
$\frac{\dot{\phi{}}^{2}}{2V}\simeq{}10^{-69}$ ,
$\epsilon_{2}\simeq{}\mathcal{O}(10^{-2})$ and the theory's
$\kappa{}\frac{\xi^{'}}{\xi^{''}}$ ratio is of order
$\mathcal{O}(10^{-3})$ while the ensuing approximation for the
further simplification of the higher-order slow-roll indices
becomes
$\frac{4\kappa^{4}\xi^{'2}V}{3\beta^{2}\xi^{''}}\simeq{}\mathcal{O}(10^{-50})$.
Therefore, not only have we managed to achieve compatibility with
the 2018 Planck constraints and slow-roll approximating conditions
but there exists substantial evidence of a blue-tilt in the tensor
spectral index $n_{\mathcal{T}}$ of primordial perturbations ,
especially in Sets 2-4.
Before closing this section let us discuss an important issue
having to do with the super-Planckian values of the scalar field
in the context of the model studied in this section. The model I
contains super-Planck values of the scalar field in order for it
to be viable, indeed for example the final value of the scalar
field at the end of inflation for this model is
$\phi_f=\frac{\sqrt{2\beta{}}(\nu{}-1)}{\kappa}$, so for set 1 we
have $\phi_f\sim 12.7\,M_p$ which is strongly super-Planckian.
This feature is due to the stringy origin of the EGB theory, and a
detection of tensor spectral index so large as in the context of
this model, could be a strong indication that the theory behind
inflation contains string corrections. It would thus be a direct
probe of string theory or even quantum gravity, at least indirect.

The super-Planck values of the scalar field are also a probe of
quantum gravity or string effects, for example the Lyth bound
\cite{Lyth:1996im} indicates that,
$$\frac{\Delta \phi}{M_p}\geq \left(\frac{r}{8} \right)^{1/2}\,N$$
where $N$ is the $e$-folding number. By taking $N\sim 60$ a future
detection of the $B$-modes with $r\sim 0.03$ would indicate that
the distance travelled by the inflaton would be $\Delta \phi\sim
3.67\,M_p$ which is super-Planckian. Even a smaller $r$ for
example $r\sim 0.003$ would yield $\Delta \phi\sim 1.1619\,M_p$.
\footnote{See also the discussion on page 16 of
\cite{LiteBIRD:2022cnt}.} Such a detection of B-modes would thus
be an indicator of some stringy origin of the underlying theory.
Although perturbation theory may not break down, however the
Trans-Planckian issues might cause theoretical inconsistencies in
the theoretical framework. We need to note though that the Lyth
bound and the Trans-Planckian issues were derived for minimally
coupled theories and not for EGB theories. Thus one must be
cautious regarding these issues, and a correct treatment would be
required to extend the bounds for EGB theories. However a strong
blue tilt is related with super-Planckian values of the scalar
field in the context of EGB and would thus be a strong hint for a
stringy origin of the underlying theory.

\subsection{ Model II: Arctangent choice for $V(\phi)$}

\begin{figure}
    \centering
    \includegraphics[width=0.45\textwidth]{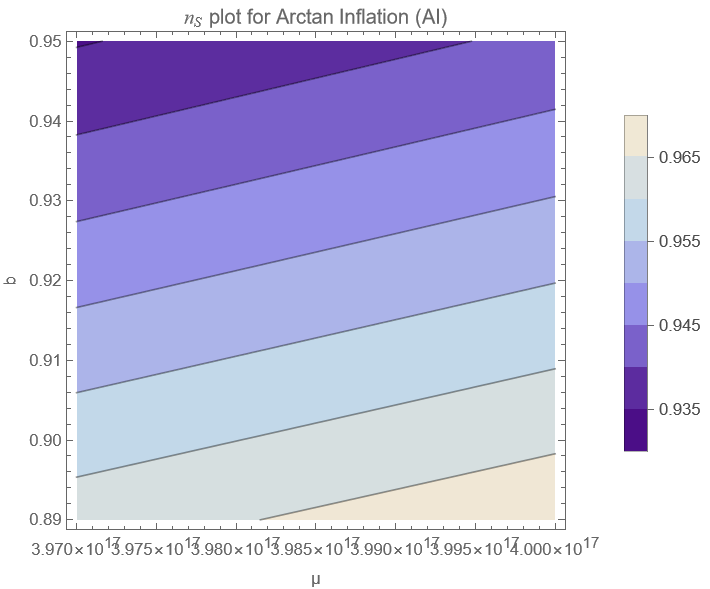}
    \includegraphics[width=0.45\textwidth]{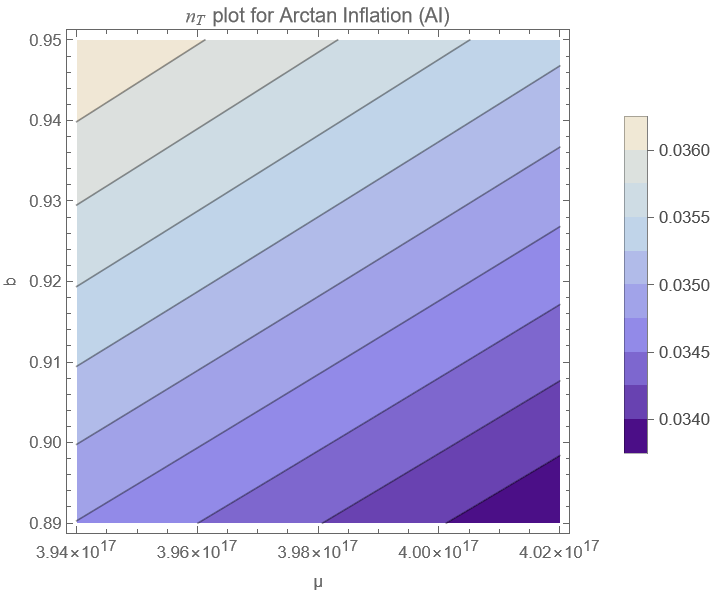}
    \includegraphics[width=0.5\linewidth]{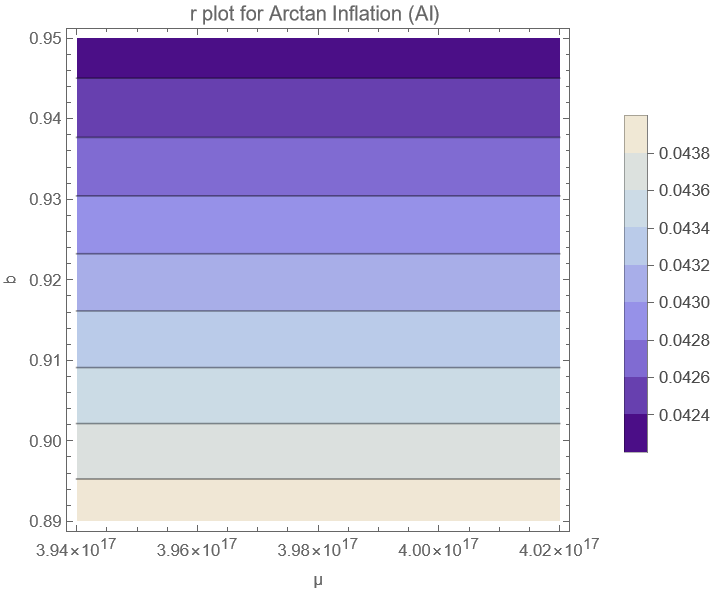}
\caption{The spectral indices for the free parameter of the theory
as well as the rescaling factor of the EBG framework. Note that in
each scenario, the indices respect the 2018 Planck constraints and
yield a positive value for the tensor spectral index, albeit no
higher than $n_{\mathcal{T}}=0.052$. Regardless, it is of
tremendous importance that both $n_{\mathcal{S}}=0.966$ and
$r<0.064$ are respected simultaneously, however the BICEP/Keck
constraints \cite{BICEP:2021xfz} are not respected for the chosen
values of the free parameters. There exist values that are also
compatible with the BICEP/Keck constraints, see for example Table
\ref{tab:2}.}
 \label{fig:10}
\end{figure}

\begin{figure}[h!]
\centering
\includegraphics[width=20pc]{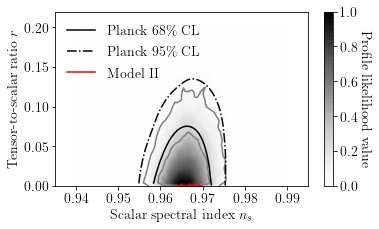}
\caption{The phenomenology of model II confronted with the Planck
2018 likelihood curves for various optimal values of the free
parameters.} \label{likelihoodmodel2}
\end{figure}

In this scenario, we examine the compatibility of inflation by an Arctan-driven potential,
\begin{gather}\vspace{0.4cm}
V(\phi)=\frac{1}{\kappa^{4}}\hspace{0.1cm}\left[1-\frac{2}{\pi{}}\arctan\left(\frac{\phi{}}{\mu{}}\right)\right] \label{eq:44}
\end{gather}
This one free-parameter model was initially considered in order to
extensively study the accuracy, with which the first and second
slow-roll order power spectra can approximate the actual power
spectrum of the inflationary cosmological perturbations. Once
again, solving Eq. (\ref{eq:15}) results in a Gauss-Bonnet
coupling constant of the form,
\begin{gather}
\xi{(\phi{},\mu{})}=-\frac{3\beta^{2}}{4}+\frac{3\beta^{2}}{4\left(1-\frac{2}{\pi}\arctan\left(\frac{\phi}{\mu}\right)\right)}
\end{gather}
It is to be highlighted that, in this instance, the slow-roll
indices obtain quite complicated expressions, due to the nature of
$\xi(\phi,\mu)$ derivatives, which we will not quote here for
brevity. Solving $\epsilon_1=1$ exactly is arduous and hardly
possible, which is precisely why the calculations were performed
up to $\mathcal{O}(3)$ in terms of $\phi{}$ , yielding the
following functional form of the $\epsilon_{1}$ slow-roll index,
\begin{gather}
    \epsilon_{1}(x)\simeq{}\frac{0.308\kappa^{2} \mu^{2}}{\beta{}}+\frac{0.574 \kappa^{2}\mu{}\phi{}}{\beta{}}+\frac{1.170\kappa^{2}\phi^{2}}{\beta{}}+\mathcal{O}\left(\phi^{3}\right)
\end{gather}
where convergence implies that the value of the ensuing ration
$\frac{\phi{}}{\mu{}}\ll{}1$ or equivalently that we're
considering a specific range of the model's free parameter.
Regardless, the scalar field value at the end of inflation reads,
\begin{gather}
\phi_f{}\simeq{}=-0.245\mu{}+0.045\sqrt{10^{18}\beta{}-97.4\mu^{2}} \\ \vspace{0.2cm}
\phi_i{}=-1.57\mu{}
\end{gather}
Before proceeding with the spectral indices plots, it is important
to consider the limitations of the model at hand. Without any
approximations to the function form of
$(\xi{(\phi{})},\epsilon_1{(\phi{}))}$ the end of inflation
provided by $\epsilon_1{}=1$ will depend solely on
$(\mu{},\beta{})$. In that scenario, cases where:
\begin{gather}
\epsilon^{max}_2\gtrsim{}\epsilon^{max}_1
\end{gather}
where by maximum we denote the slow-roll values at $\phi_{max}$,
then the slow-roll framework breaks down and cannot accurately
describe the dynamics of inflation.  Due to the odd symmetry of
the potential about $x\rightarrow-x$ translations, negative values
as above are accepted, resulting in $V(\phi{})>0$ altogether.
Before proceeding with the plots, it will be made clear that for
illustrative and qualitative purposes the range of the parameters
are set to best display the characteristics of each index. As
such, in the simultaneous compatibility check it will be evident
that the parameter range contributes insignificantly to the end
results. Therefore, the spectral indices for the arctangent
framework are displayed in Fig. \ref{fig:10}.
It is to be noted, that both tensor and scalar spectral indices
exhibit values that correspond to a blue-tilted inflationary
regime. Moreover, by examining the compatibility with the rest of
the spectral indices we gathered the viable phenomenologies in
Table \ref{tab:2}.
\begin{table}
    \centering
    \begin{tabular}{|c|c|c|c|c|c|} 
        \hline
        &   $\mu{}$&$\beta{}$&$n_{\mathcal{S}}$& $n_{\mathcal{T}}$& $r$\\ 
        \hline
        Set 1)&   $3.8\cdot{}10^{18}$&0.89&0.966& 0.0348& $7.6\cdot{}10^{-17}$\\ \hline  
        Set 2)&   $3.475\cdot{}10^{17}$&0.7&0.966& 0.035& $9\cdot{}10^{-18}$\\ \hline  
        Set 3)&   $2.672\cdot{}10^{17}$&0.6&0.966& 0.035& $1.4\cdot{}10^{-17}$\\ \hline  
         Set 4)&   1.32$\cdot{}10^{17}$&0.1&0.966& 0.034& $1.33\cdot{}10^{-18}$\\\hline
    \end{tabular}
    \caption{Validity of a blue-tilted $n_{\mathcal{T}}$ inflationary regime
for an Arctangent-driven potential, accounting for various values
of the rescaling and the model's free parameter.}
    \label{tab:2}
\end{table}
Despite the spectral yield of tensor perturbations not being as
significantly large as in the power-law scenario, it still
consists of a blue-tilt and a step forward in terms of
progression. We may finalize the analysis of this model only after
providing the slow-roll approximation accordance with the selected
values, as listed in Table \ref{tab:3}. As it can be seen, both
the Planck and the BICEP/Keck constraints are respected for this
model too.
\begin{table}
    \centering
    \begin{tabular}{|c|c|c|c|c|c|l|l|} 
        \hline
        &   $\mu{}$&$\beta{}$&$\epsilon_{1}$& $\frac{\dot{\phi}^{2}}{2V}$ & $\epsilon_{2}$& $\kappa{}\frac{\xi^{'}}{\xi^{''}}$& $\frac{4\kappa^{4}\xi^{'2}V}{3\beta^{2}\xi^{''}}$\\ 
        \hline
        Set 1)&   $3.8\cdot{}10^{18}$&0.89&$\mathcal{O}(10^{1})$& $\mathcal{O}(10^{-1})$& $\mathcal{O}(10^{-1})$& $\mathcal{O}(10^{1})$&$\mathcal{O}(10^{-2})$\\ \hline  
        Set 2)&   $3.475\cdot{}10^{17}$&0.7&$\mathcal{O}(10^{-2})$& $\mathcal{O}(10^{-3})$& $\mathcal{O}(10^{-1})$& $\mathcal{O}(10^{-1})$&$\mathcal{O}(10^{-2})$\\ \hline  
        Set 3)&   $2.672\cdot{}10^{17}$&0.6&$\mathcal{O}(10^{-3})$& $\mathcal{O}(10^{-3})$& $\mathcal{O}(10^{-1})$& $\mathcal{O}(10^{-2})$&$\mathcal{O}(10^{-2})$\\ \hline  
         Set 4)&   1.32$\cdot{}10^{17}$&0.1&$\mathcal{O}(10^{-2})$& $\mathcal{O}(10^{-3})$& $\mathcal{O}(10^{-1})$& $\mathcal{O}(10^{-2})$&$\mathcal{O}(10^{-2})$\\\hline
    \end{tabular}
    \caption{Validity of the slow-roll approximations of equations (4.8) and
(4.16)-(4.18). Notice how only one case scenario does not meet the
necessary prerequisites.}
    \label{tab:3}
\end{table}
Parameter set 1 is not in accordance with the slow-roll demands as
it is way out of the slow-roll range $\epsilon_{1}>1$, hence its
omittance from compatible sets. In addition, in Fig.
\ref{likelihoodmodel2} we confront model II with the Planck 2018
likelihood curves for various suitably chosen values of the free
parameters. As it can be seen, the model is nicely fitted in the
Planck likelihood curves and the model is phenomenologically
fitted in the sweet spot of the Planck data.

\subsection{Model III: Double-Well Inflation for $V(\phi{})$}

\begin{figure}
    \centering
    \includegraphics[width=0.45\textwidth]{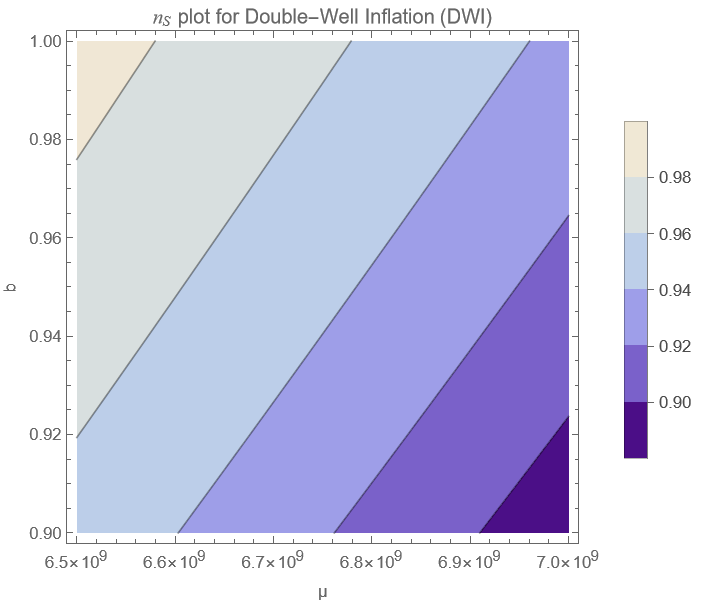}
    \includegraphics[width=0.45\textwidth]{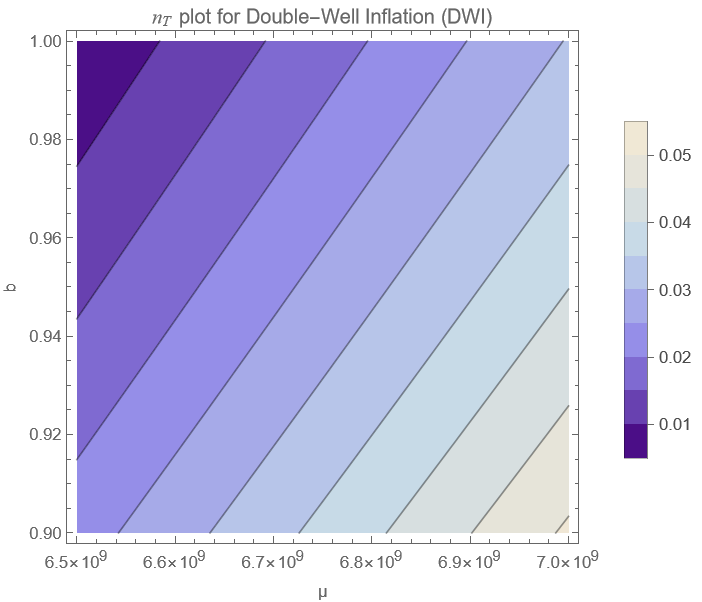}
    \includegraphics[width=0.5\linewidth]{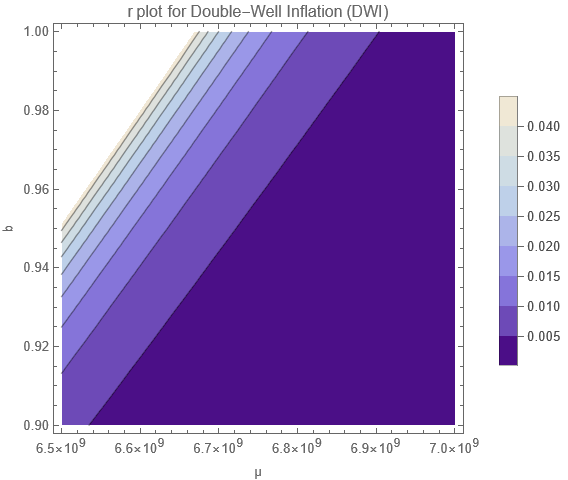}
\caption{The above pictures portray the qualitative and
quantitative trends of the spectral indices of primordial
perturbations, for a range of values within the margins of 2018
Planck compatibility and, possibly, blue-tilt of
$n_{\mathcal{T}}$. It is important to note that the values of the
tensor-to-scalar ratio of the perturbation's amplitudes have been
rescaled as $r'\rightarrow{}10^{14}r$ in order to display its
features in a palpable fashion, whereas the values of $\mu$ are of
the order $10^{9}\cdot{}GeV$.}
  \label{fig:15}
\end{figure}

\begin{figure}
    \centering
    \includegraphics[width=0.45\textwidth]{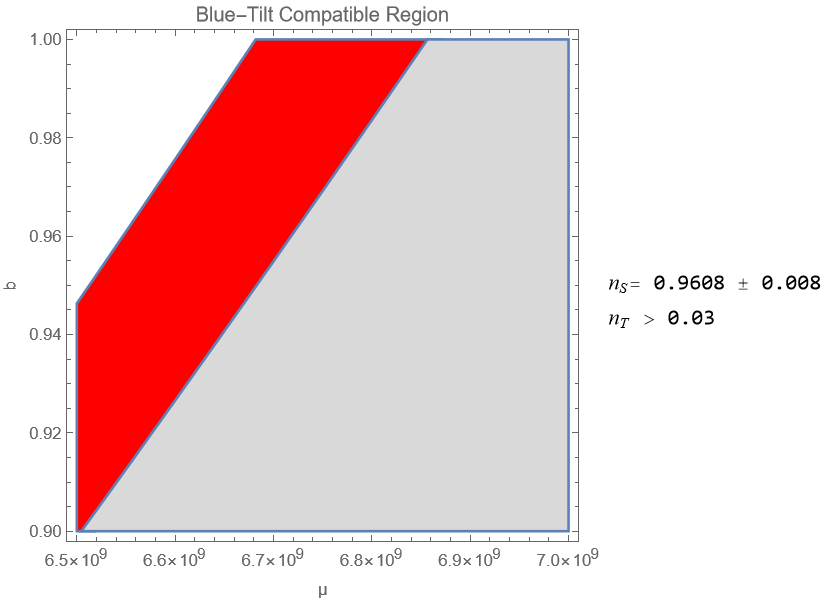}
   \caption{Blue tilt compatibility of the $n_{\mathcal{T}}$ , with respect to
the 2018 Planck constraints regarding the allowed range of values
for $n_{\mathcal{S}}$ and tensor-to-scalar ratio $r$}
    \label{fig:Compatibility}
\end{figure}

\begin{figure}[h!]
\centering
\includegraphics[width=20pc]{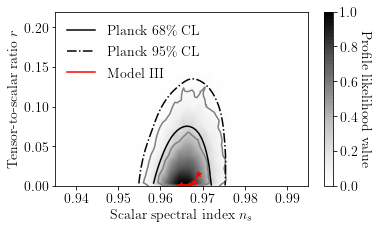}
\caption{The phenomenology of model III confronted with the Planck
2018 likelihood curves for various optimal values of the free
parameters.} \label{likelihoodmodel3}
\end{figure}

In this model, we examine another one-parameter potential , the
''Double-Well'' or ''Mexican hat'' case which was first introduced
by Goldstone as a candidate for symmetry breaking. However, in the
inflationary framework it is employed to investigate the formation
and structure of topological defects in spacetime. The potential
reads,
\begin{gather}
V(\phi{})=\frac{1}{k^4}\left[\left(\frac{\phi}{\mu}\right)^2-1\right]^2
\end{gather}
with $[\mu{}]=[m]$ and its respective non-minimally coupling
function $\xi{(\phi)}$ will be deduced from (\ref{eq:15}),
\begin{gather}
\xi{(\phi)}=\frac{3 b^2}{4 \left[\left(\frac{\phi}{\mu}\right)^2-1\right]^2}
\end{gather}
Regarding the presentation of the slow-roll indices despite their lengthy expressions, as per usual, we've obtained,
\begin{align}
    \epsilon_{1}&\simeq{}\frac{\kappa^2\phi^2\left(\phi^2-\mu^2\right)^2}{2\beta\left(5\phi^2+\mu^2\right)^2}
 \\ \notag \\
    \epsilon_{2}&\simeq{}-\frac{\left(\kappa^2\phi^2\left(\phi^2-\mu^2\right)^2+2\beta\left(5\phi^4+8\phi^2\mu^2-\mu^4\right)\right)}{2\beta\left(5\phi^2+\mu^2\right)^2
} \\ \notag \\
\epsilon_{3}&\simeq{}0 \\ \notag \\
\epsilon_{4}&\simeq{}\frac{24 \beta \phi^2 \left(5 \phi^6 + 7 \phi^4 \mu^2 + 11 \phi^2 \mu^4 + \mu^6\right)}{\left(\phi^2 + \mu^2\right) \left(5 \phi^2 + \mu^2\right) \left(\kappa^2 \left(\phi^2 - \mu^2\right)^2 \left(\phi^2 + \mu^2\right) + 24 \beta \phi^2 \left(5 \phi^2 + \mu^2\right)\right)} \\ \notag \\
\epsilon_{5}&\simeq{}-\frac{2 \phi^2}{\phi^2 + \mu^2} \\ \notag \\
\epsilon_{6}&\simeq{}\frac{4 \mu^2 \phi^2 \left(\phi^2 - \mu^2\right)}{\left(\mu^2 + \phi^2\right) \left(\mu^2 + 5 \phi^2\right)^2}
\end{align}
thereby proceed to evaluate the initial modes entering the Hubble
horizon by finding the end of inflation through $\epsilon_{1}=1$
up to $\mathcal{O}(3)$  this time then we find,
\begin{gather}
\phi_{i}=\frac{
    \left(9 \beta^{13/2} \left(\frac{1}{\kappa^2}\right)^{3/2} \left(6 \beta + \kappa^2 \mu^2\right) + \zeta\left(\mu{},\beta{}\right)\right)^{2/3} - \sqrt[3]{2} e^{40} \beta^4 \mu^2
}{
    3 \sqrt[6]{2} e^{20} \beta^2 \sqrt[3]{9 \beta^{13/2} \left(\frac{1}{\kappa^2}\right)^{3/2} \left(6 \beta + \kappa^2 \mu^2\right) + \zeta\left(\mu{},\beta{}\right)}
} \\ \notag \\
\zeta\left(\mu{},\beta{}\right)=\frac{
    \beta^{12} \left(2916 \beta^3 + 972 \beta^2 \kappa^2 \mu^2 + 81 \beta \kappa^4 \mu^4 + 2 e^{120} \kappa^6 \mu^6\right)
}{
    \kappa^6
}
\end{gather}
Taking into consideration that the slow-roll regime for this
potential is valid only for super-Planckian values of the inflaton
field, the exact same problem that was encountered in the
arctangent case, we demand the very weak (by cosmological
standards) condition that the value of $\epsilon_{2}(0)<1$ , in
other words ,
\begin{gather}
k\cdot\mu>2\sqrt{2} \Rightarrow \mu{}>2\sqrt{2}\cdot\mathcal{M}_{Pl}
\end{gather}
With the constraints in mind, evaluation of the spectral indices
at the first horizon crossing yields the plots of Fig.
\ref{fig:15}.
Therefore, since has been shown that a blue-tilted inflation takes
place in the super-Planckian region of values of the scalar field
what remains is to validate the slow-roll approximations for this
theory. Much like in every other scenario investigated so far, we
will consider four different data sets yielding a blue-tilt and
construct a table containing the orders of magnitude for each
slow-roll approximating condition. All of the results are present
in Table \ref{tab:4}.
\begin{table}
    \centering
    \begin{tabular}{|c|c|c|c|c|c|l|l|} 
        \hline
        &   $\mu{}(10^{9}\cdot{}GeV)$&$\beta{}$&$\epsilon_{1}$& $\frac{\dot{\phi}^{2}}{2V}$ & $\epsilon_{2}$& $\kappa{}\frac{\xi^{'}}{\xi^{''}}$& $\frac{4\kappa^{4}\xi^{'2}V}{3\beta^{2}\xi^{''}}$\\ 
        \hline
        Set 1)&   6.5&0.92&$\mathcal{O}(10^{-21})$& $\mathcal{O}(10^{-22})$& $\mathcal{O}(10^{-1})$& $\mathcal{O}(10^{-11})$&$\mathcal{O}(10^{-1})$\\ \hline  
        Set 2)&   6.6&0.94&$\mathcal{O}(10^{-21})$& $\mathcal{O}(10^{-21})$& $\mathcal{O}(10^{-1})$& $\mathcal{O}(10^{-11})$&$\mathcal{O}(10^{-1})$\\ \hline  
        Set 3)&   6.7&0.98&$\mathcal{O}(10^{-21})$& $\mathcal{O}(10^{-22})$& $\mathcal{O}(10^{-1})$& $\mathcal{O}(10^{-11})$&$\mathcal{O}(10^{-1})$\\ \hline  
         Set 4)&   6.9&1&$\mathcal{O}(10^{-21})$& $\mathcal{O}(10^{-21})$& $\mathcal{O}(10^{-1})$& $\mathcal{O}(10^{-11})$&$\mathcal{O}(10^{-2})$\\\hline
    \end{tabular}
    \caption{Validity of the slow-roll approximations for the Double-Well Inflation (DWI)}
    \label{tab:4}
\end{table}
On a last note, we also add the respective values of the spectral
indices of the primordial perturbation to get a qualitative
overview of our results, presented in Table \ref{tab:5}. As it can
be seen, both the Planck and the BICEP/Keck constraints
\cite{BICEP:2021xfz} are respected for this model too.
\begin{table}
    \centering
    \begin{tabular}{|c|c|c|c|c|c|} 
        \hline
        &   $\mu{}(10^{9}\cdot{}GeV)$&$\beta{}$&$n_{\mathcal{S}}$& $n_{\mathcal{T}}$& $r(10^{-14})$\\ 
        \hline
        Set 1)&   6.5&0.92&0.966& 0.02& 0.015\\ \hline  
        Set 2)&   6.6&0.94&0.964& 0.025& 0.0012\\ \hline  
        Set 3)&   6.7&0.98&0.968& 0.03& 0.001\\ \hline  
         Set 4)&   6.9&1&0.965& 0.051& 0.005\\\hline
    \end{tabular}
    \caption{Values of the spectral indices under the rescaled EGB framework,
for the Double-Well Inflation. Note how blue-tilt is achieved in
each case, with simultaneous compatibility with the 2018 Planck
constraints regarding $n_{\mathcal{S}}\simeq{}0.966$ as well as
$r<0.064$. Thus both the Planck and the BICEP/Keck constraints
\cite{BICEP:2021xfz} are respected for this model too.}
    \label{tab:5}
\end{table}
A brief overview of the results may be synopsized in the graph of
Fig. \ref{fig:Compatibility}, where the tensor spectral index may
exhibit a blue-tilt that is in accordance with the imposed Planck
constraints ,
Finally, in Fig. \ref{likelihoodmodel3} we confront model III with
the Planck 2018 likelihood curves for various suitably chosen
values of the free parameters. In this case too, the model is
nicely fitted in the Planck likelihood curves nd some values of
the free parameters put the model predictions in the sweet spot of
the Planck data.

\subsection{Model IV: Natural Inflation model for $V(\phi)$ (NI)}

\begin{figure}
    \centering
    \includegraphics[width=0.45\textwidth]{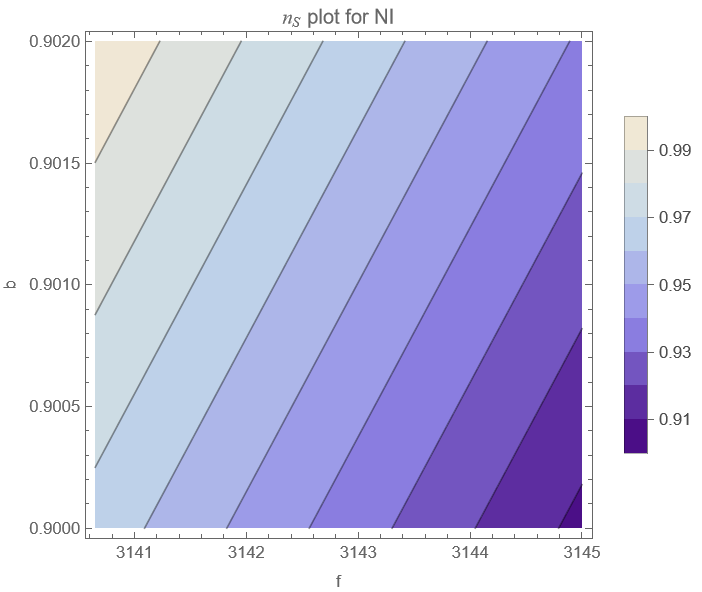}
    \includegraphics[width=0.45\textwidth]{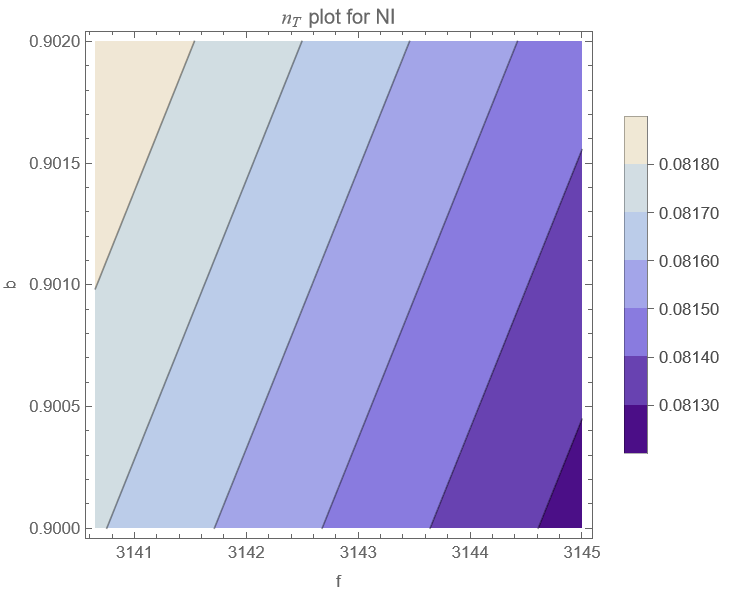}
    \includegraphics[width=0.5\linewidth]{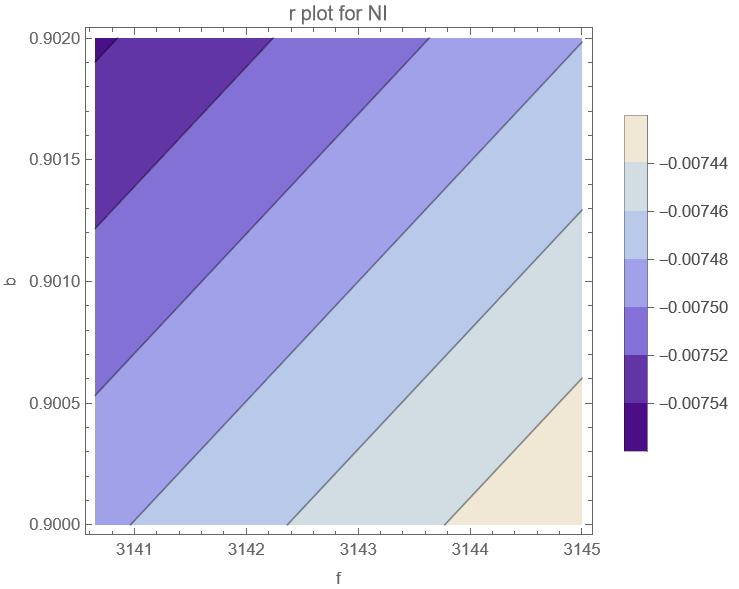}
\caption{The qualitative and quantitative trends of the spectral
indices of primordial perturbations, for a range of values within
the margins of 2018 Planck compatibility and, possibly, blue-tilt
of $n_{\mathcal{T}}$.}
 \label{fig:18}
\end{figure}

\begin{figure}[h!]
\centering
\includegraphics[width=20pc]{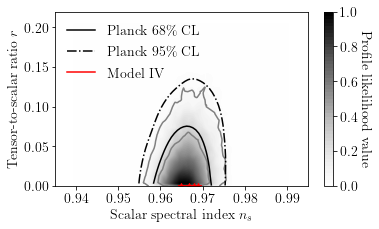}
\caption{The phenomenology of model IV confronted with the Planck
2018 likelihood curves for various optimal values of the free
parameters.} \label{likelihoodmodel4}
\end{figure}

Natural inflation was first proposed as an attempt to solve the
fine-tuning problem of inflation, and it is closely related to
axion fields \cite{Caputo:2024oqc,Kuster:2008zz,Marsh:2015xka}. In
particular, in order to obtain sufficient inflation and the
correct normalization for the microwave background anisotropies,
the potential $V(\phi)$ of the inflaton must be sufficiently flat.
Since we are only interested in verifying whether or not this
model can reproduce blue-tilted spectral indices, the main focus
will be on the qualitative analysis of the slow-roll parameters
that derive from this potential,
\begin{gather}
    V(\phi)=\frac{1}{k^4}\left[1+\cos{\left(\frac{\phi}{f}\right)}\right]
\end{gather}
in which case the parameter $f$ , with $[f]=[m]$ represents the
a-priori unknown, symmetry breaking scale. With that in mind , we
make use of Eq. (\ref{eq:15}) to derive the non-minimally coupling
function to the theory,
\begin{gather}
\xi(\phi)=\frac{3 b^2}{4 \left(\cos \left(\frac{\phi}{f}\right)+\gamma  k^4+1\right)}
\end{gather}
where the introduction of the $\gamma>0$ term stems from the need
to avoid divergences in the coupling function. After computing the
slow-roll indices for this coupling function we get,
\begin{align}
    \epsilon_{1}&\simeq{}\frac{f^2 \kappa^2 \sin^2\left(\frac{\phi}{f}\right)}{2 \beta \left(\cos\left(\frac{\phi}{f}\right) - 2\right)^2} \\ \notag \\
    \epsilon_{2}&\simeq{}\frac{-4\beta-f^2\kappa^2+8\beta \cos\left(\frac{\phi}{f}\right)+f^2\kappa^2\cos\left(\frac{2 \phi}{f}\right)
}{4\beta\left(-2+\cos\left(\frac{\phi}{f}\right)\right)^2} \\ \notag \\
    \epsilon_{3}&\simeq{}0 \\ \notag \\
\end{align}
\begin{align}
    \epsilon_{4}&\simeq{}\frac{
    3 \beta \left[-9 + 4 \cos\left(\frac{\phi}{f}\right) + \cos\left(\frac{2 \phi}{f}\right)\right] \sin^2\left(\frac{\phi}{2 f}\right)
}{\left(-2 + \cos\left(\frac{\phi}{f}\right)\right) \left(15 \beta + 4 f^2 \kappa^2 + \left(-18 \beta + 4 f^2 \kappa^2\right) \cos\left(\frac{\phi}{f}\right) + 3 \beta \cos\left(\frac{2 \phi}{f}\right)\right)} \\ \notag \\
    \epsilon_{5}&\simeq{}-\sin^{2}\left(\frac{\phi{}}{2 f}\right) \\ \notag \\
    \epsilon_{6}&\simeq{}-\frac{\sin^2\left(\frac{\phi}{f}\right)}{2 \left(\cos\left(\frac{\phi}{f}\right) - 2\right)^2}
\end{align}
all of which constitute well-defined functions with no
irregularities in their expressions, a feature that will prove
invaluable later on. Having obtained the functional forms of the
slow-roll indices, the end of inflation is easily obtained if one
considers $\epsilon_{1}(\phi_{f})=1$. This poses a transcendental
equation that has no algebraic solution, but given that in the
Planckian scale the ratio $x=\frac{\phi{}}{f}\\{}1$ it is
permissible to express the first slow-roll index with its leading
order in its respective Taylor series expansion, yielding,
\begin{gather}
\epsilon_{1}(\phi{})\simeq{}\frac{\kappa^2 \phi^2}{2 \beta} - \frac{2 \kappa^2 \phi^4}{3 \beta f^2}
+\mathcal{O}(\phi{})^{6}
\end{gather}
the resulting solutions upon substituting the approximated functional form become,
\begin{gather}
    \phi_{f}=\pm{}\sqrt{\frac{\sqrt{3} f \sqrt{3 f^2 \kappa^4 - 32 \beta \kappa^2}}{8 \kappa^2} + \frac{3 f^2}{8}}
\hspace{0.2cm}\text{or}\pm{}\frac{\sqrt{3 f^2 - \frac{\sqrt{3} f \sqrt{\kappa^2 \left(-32 \beta + 3 f^2 \kappa^2\right)}}{\kappa^2}}}{2 \sqrt{2}}
\end{gather}
meanwhile, the value of the field at the first horizon crossing
will be made following a simple expansion of $\xi^{'}$ around
$\phi{}=0$,
\begin{gather}
    \xi^{'}(\phi{})=\frac{3 \beta^2 \phi}{8f^2 } + \mathcal{O}(x^{3})
\end{gather}
where obviously even powers of $x$ vanish in the series as the
derivative of the scalar coupling function, being an even function
itself, will be odd overall. This leads to a value of the scalar
field at the first horizon crossing that reads,
\begin{gather}
    \phi_{i}=\frac{\sqrt{2\beta{}}}{e^{N}\kappa{}}
\end{gather}
with $N=60$ being the number of e-folds at the end of inflation.
Substituting into the spectral indices we immediately observe that
$r\rightarrow{}0$ , since it is on the scale of $10^{-40}$ safely
within the margins of the Planck observational data. These
approximations have been handled with extreme diligence to ensure
the qualitative behavior of the spectral indices. In hindsight,
one could obtain very similar final results by considering even
lower leading-order Taylor series due to the great convergence
satisfied by the Planckian values of the a-priori unknown scale
$f$. Finally, the graphs of Fig. (\ref{fig:18}) encapsulate all
there is to know about the spectral indices at the first horizon
crossing,
The plots of Fig. \ref{fig:18} portray the qualitative and
quantitative trends of the spectral indices of primordial
perturbations, for a range of values within the margins of 2018
Planck compatibility and, possibly, blue-tilt of
$n_{\mathcal{T}}$. It is important to note that the values of the
tensor-to-scalar ratio of the perturbation's amplitudes have been
rescaled as $r'\rightarrow{}10^{14}r$ in order to display its
features in a palpable fashion, whereas the values of the symmetry
breaking scales are of the order of $10^{18}\text{ GeV}$. Thereby
affirming the Planckian permission of
observational-constraint-respecting inflation and hinting at a
possible blue tilt of the tensor spectral index. Also, it is to be
noted that the tensor-to-scalar ratio pertains to values whose
magnitude respects the 2018 Planck constraints imposing $r<0.064$
and in addition the BICEP/Keck constraints \cite{BICEP:2021xfz}
are respected too. Lastly, we will concern ourselves with the
validity of the slow-roll approximations for selected sets of
blue-tilt satisfying values of the free parameters
$\left(f,\beta{}\right)$. Our results for viable cases listed in
Table \ref{tab:6}, are accompanied by the validity of the
approximations which are estimated in orders of magnitude as seen
in Table \ref{tab:7}.
\begin{table}
    \centering
    \begin{tabular}{|c|c|c|c|c|c|} 
        \hline
        &   $f(10^{18}GeV)$&$\beta{}$&$n_{\mathcal{S}}$& $n_{\mathcal{T}}(10^{-30})$& $r(10^{-18})$\\ 
        \hline
        Set 1)&   3141&0.905&0.966& 8.2& 7.65\\ \hline  
        Set 2)&   3142&0.91&0.962& 8.24& 7.75\\ \hline  
        Set 3)&   3143&0.9105&0.967& 8.5& 7.89\\ \hline  
         Set 4)&   3144&0.93&0.966& 9.13& 8.08\\\hline
    \end{tabular}
    \caption{Viable Scenarios for Natural inflation.}
    \label{tab:6}
\end{table}
\begin{table}
    \centering
    \begin{tabular}{|c|c|c|c|c|c|l|l|} 
        \hline
        &   $f(10^{18}GeV)$&$\beta{}$&$\epsilon_{1}$& $\frac{\dot{\phi}^{2}}{2V}$ & $\epsilon_{2}$& $\kappa{}\frac{\xi^{'}}{\xi^{''}}$& $\frac{4\kappa^{4}\xi^{'2}V}{3\beta^{2}\xi^{''}}$\\ 
        \hline
        Set 1)&   3141&0.905
&$\mathcal{O}(10^{-80})$& $\mathcal{O}(10^{-84})$& $\mathcal{O}(10^{-1})$& $\mathcal{O}(10^{-34})$&$\mathcal{O}(10^{-53})$\\ \hline  
        Set 2)&   3142&0.91
&$\mathcal{O}(10^{-80})$& $\mathcal{O}(10^{-84})$& $\mathcal{O}(10^{-1})$& $\mathcal{O}(10^{-34})$&$\mathcal{O}(10^{-53})$\\ \hline  
        Set 3)&   3143&0.9105
&$\mathcal{O}(10^{-80})$& $\mathcal{O}(10^{-84})$& $\mathcal{O}(10^{-1})$& $\mathcal{O}(10^{-31})$&$\mathcal{O}(10^{-53})$\\ \hline  
         Set 4)&   3144&0.93&$\mathcal{O}(10^{-80})$& $\mathcal{O}(10^{-84})$& $\mathcal{O}(10^{-1})$& $\mathcal{O}(10^{-34})$&$\mathcal{O}(10^{-53})$\\\hline
    \end{tabular}
    \caption{Validity of the slow-roll approximations for the Natural inflation scenario. Notice how only one case scenario does not meet the necessary prerequisites.}
    \label{tab:7}
\end{table}
Therefore, all of our data sets that satisfy the 2018 Planck
constraints are also well approximated within what is permissible
from the slow-roll conditions. In addition, in Fig.
\ref{likelihoodmodel4} we confront model IV with the Planck 2018
likelihood curves, again for various suitably chosen values of the
free parameters. As it can be seen in this case too, the model is
nicely fitted in the sweet spot of the Planck likelihood curves.

\subsection{Model V: Logarithmic Potential Inflation model for $V(\phi)$ (LPI)}

In this case we consider the potential of Logarithmic Potential
Inflation (LPI) theory, namely,
\begin{equation}\label{logpotential}
    V(\phi) = M^4 \left( \frac{\phi}{\phi_0} \right)^p \left( \ln \frac{\phi}{\phi_0} \right)^q
\end{equation}
The choices ($p=4,$ $q=1)$, ($p=1$, $q=2$), ($p=4$, $q=3$) match
different Yang-Mills composite models. In the LPI models $M$ and
$\phi_0$ are related to a mass scale of the underlying theory,
similarly here, $\phi_0$ sets the scale at which inflation can
occur and is permitted to take either super-Planckian or
sub-Planckian values. The parameter $p$ is not bounded to any
restrictions, while the parameter $q$ has to be an integer (and in
some cases, an even integer) to ensure a well-defined potential.
Solving Eq. (\ref{eq:15}), we find the coupling function to be,
\begin{equation}
   \xi(\phi) = \frac{3 \left( \frac{\phi}{\phi_0} \right)^{-P} \ln \left( \frac{\phi}{\phi_0} \right)^{-q}}{4 M^4 \kappa^4\beta^2}
\end{equation}
and thus, we can proceed to the slow-roll analysis of the potential.

\begin{figure}
    \centering
    \includegraphics[width=0.5\linewidth]{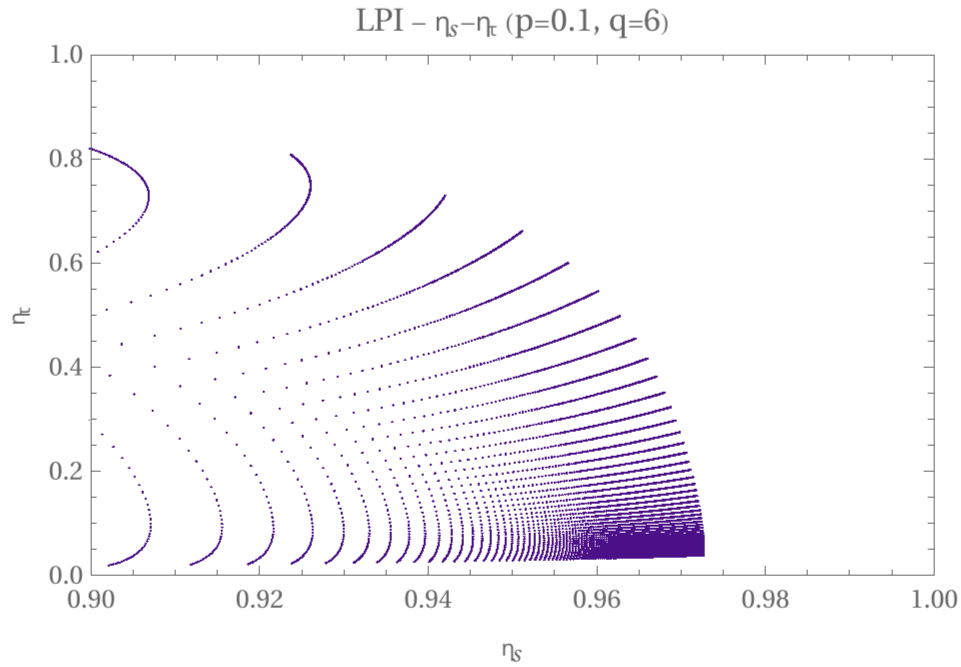}
    \caption{List plot of $n_{\mathcal{S}}-n_{\mathcal{T}}$, with $p=0.1$, portrays a dense region of points viable to the 2018 Planck data constraints with a significant blue-tilt.}
    \label{fig:LPIextraplots1}
\end{figure}

\begin{figure}
    \centering
    \includegraphics[width=0.45\textwidth]{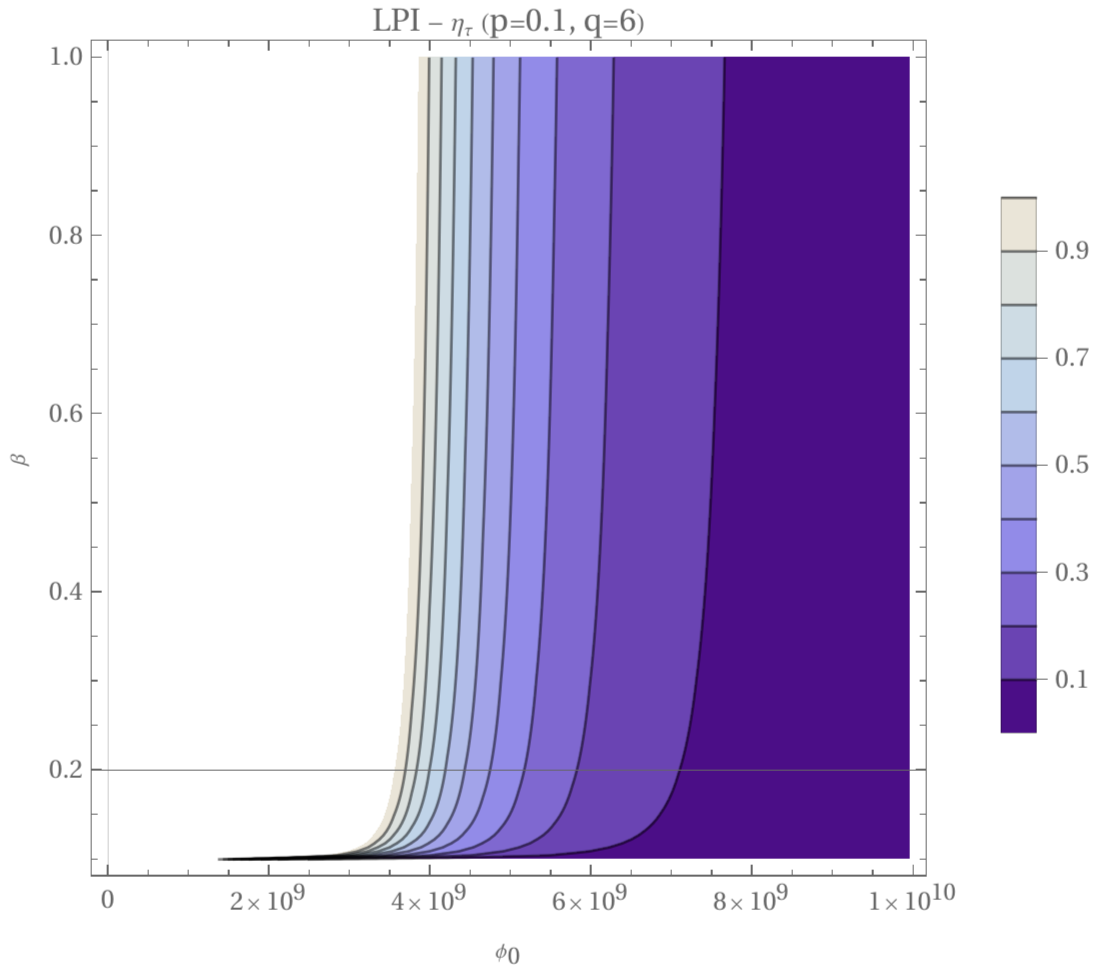}
    \hfill
    \includegraphics[width=0.45\textwidth]{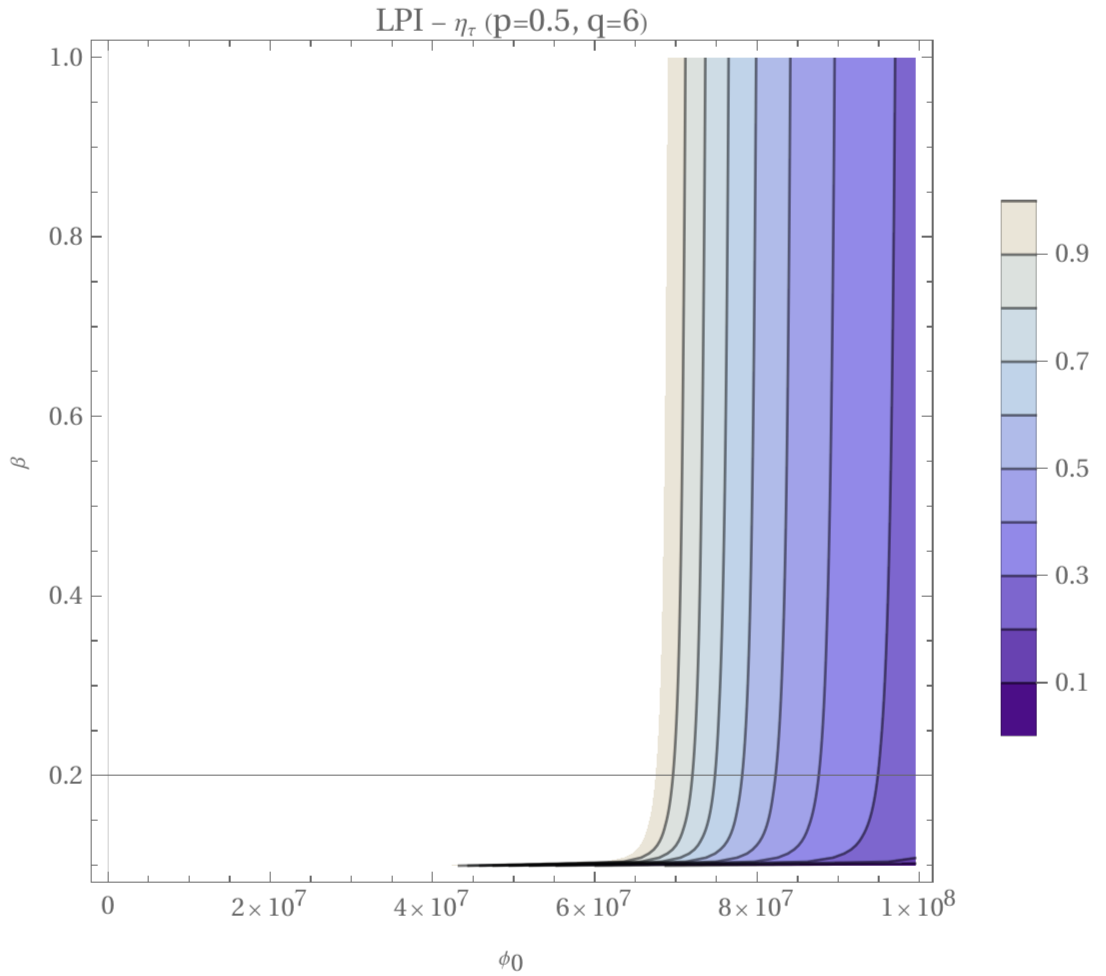}
    \vspace*{1cm}
    \includegraphics[width=0.45\textwidth]{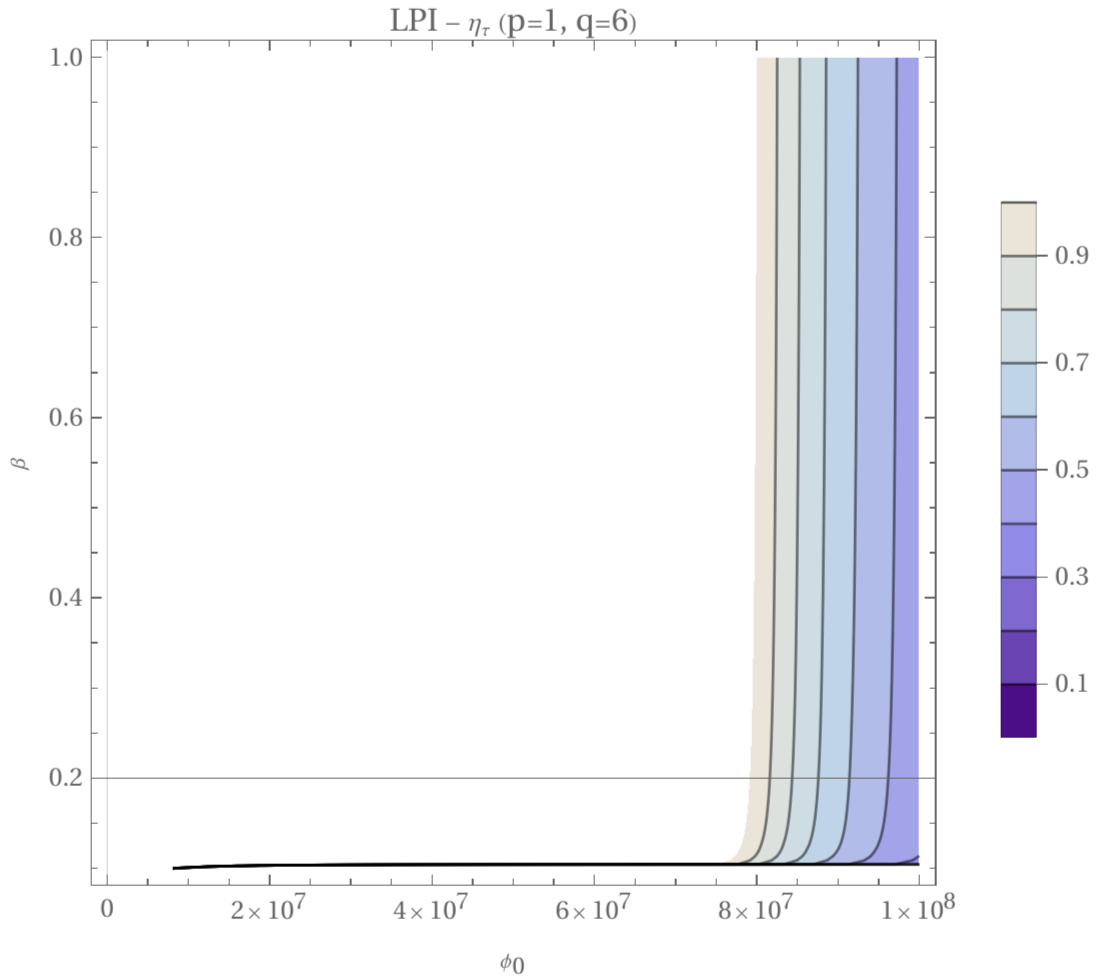}
    \caption{
        This figure displays a comparison between the $n_{\mathcal{T}}$ contours for three different cases:
        (a) Contour plot of tensor spectral index for $p=0.1$,
        (b) Contour plot of tensor spectral index for $p=0.5$, and
        (c) Contour plot of tensor spectral index for $p=1$.
        Notice how $n_{\mathcal{T}}$ becomes relevant to our range after the anomalous region, which is visible in $n_{\mathcal{S}}$ contours.
    }
    \label{fig:LPIextraplots2}
\end{figure}

If, for convenience, we set $\frac{\phi}{\phi_0}=x$, $p(p+1)=P$, $q(q+1)=Q$, the slow-roll indices for the LPI scenario can take the form,
\begin{gather}
   \epsilon_1\simeq\frac{\kappa^2 \phi^2 }{2 \beta }\frac{\log \left(x\right)^2 \left( q + p \log \left( x \right) \right)^2}{\left( Q + \log \left( x \right) \left( q + 2 p q + P \log \left( x \right) \right) \right)}\\ \notag\\
   \epsilon_2 \simeq - \frac{q Q + a_1 \ln(x) + a2 \ln^2(x) + a_3 \ln^3(x) + a_4 \ln^4(x)}{\left( Q + (2p + 1) q \ln(x) + P \ln^2(x) \right)^2}\\ \notag\\
   \epsilon_3 \simeq 0\\ \notag\\
   \epsilon_5 \simeq - \frac{\left( p \ln(x) + q \right)^2}{2 \beta^4 \left( \log(x) \right) \left( (2 p + 1) q + P \log(x) + Q \right)}\\ \notag\\
   \epsilon_6 \simeq -\left( q + p \ln(x) \right)^2 \frac{Q^2 + b_1 \ln(x) + b_2 \ln(x)^2 + b_3 \ln(x)^3 + b_4 \ln(x)^4}{2 \beta^4 \left( Q + q \left(1 + 2 p\right) \ln(x) + P \ln(x)^2 \right)^3}\\ \notag\\
\end{gather}
where $a_i$ and $b_i$ are functions of $p$ and $q$, that are
defined as follows,
    \begin{align*}
        a_1 &= Q(2 p + q) \\
        a_2 &= P + q \left( 3p + 1 + \frac{\kappa^2 \phi^2}{2 \beta} \right)\\
        a_3 &= 2 q p \left( \frac{3p}{2} + 1 + \frac{\kappa^2 \phi^2}{2 \beta} \right) \\
        a_4 &=  p \left( P + p \frac{\kappa^2 \phi^2}{2 \beta} \right) \\
    \end{align*}
    \begin{align*}
        b_1 &= 2 Q q \left( 1 + 2 p \right) \\
        b_2 &= q \left( 2 P (1 + 3 q) + q \left(1 - \frac{\kappa^2 \phi^2}{2 \beta}\right) \right) \\
        b_3 &= 2 q \left( P (1 + 2 p) - \frac{\kappa^2 \phi^2}{2 \beta} \right) \\
        b_4 &= P^2 - \frac{\kappa^2 \phi^2}{2 \beta}
    \end{align*}
The slow-roll index $\epsilon_4$ has a rather complicated
expression and is not included as it extends to the 6th order in
$\ln(x)$, with higher complexity in the coefficients. Using the
Eqs. (\ref{eq:28}-\ref{eq:30}) we can proceed to the scalar
spectral index $n_{\mathcal{S}}$, the tensor spectral index
$n_{\mathcal{T}}$ and the tensor-to-spectral ratio $r$. If we
further define $Q_p = q \left(1 + 2 p \right)$, $Q_\beta =
\frac{q^2}{2} + Q \, \beta^4$ and $P_\beta = \frac{p^2}{2} + P \,
\beta^4$, we can write:
\begin{gather}
    n_{\mathcal{S}} = 1 - 4\epsilon_1 - 2\epsilon_2 - 2\epsilon_4\\ \notag\\
    n_{\mathcal{T}} = \frac{\left( q + p \, \ln(x) \right)^2 \, Q^2 + 2 c_1 \, \ln(x) + c_2 \, \ln(x)^2 + 2 c_3 \, \ln(x)^3 + c_4 \, \ln(x)^4}{\beta^4 \, \left( Q + Q_p \ln(x) + P \, \ln(x)^2 \right)^3}\\ \notag\\
    r = 16\epsilon_1
\end{gather}
where the extra functions $c_i$ are described by,
    \begin{align*}
        c_1 &= 2 Q \, Q_p\\
        c_2 &= 2 p \, q \, (1 + 3 q) + q^2 - Q_\beta \frac{\kappa^2 \phi^2}{2 \beta}\\
    \end{align*}
    \begin{align*}
        c_3 &= P Q_p - \left(p q + \beta^4 Q_p \right) \frac{\kappa^2 \phi^2}{2 \beta} \\
        c_4 &=  P^2 - 2 P_\beta \frac{\kappa^2 \phi^2}{2 \beta}
    \end{align*}
Since $n_{\mathcal{S}}$ depends on $\epsilon_4$, the full
expression is also not displayed for brevity. Unfortunately, due
to complex form of the slow-roll parameters, the condition
$\epsilon_1=1$ at the end of inflation and Eqs. (\ref{eq:32})
cannot be solved analytically, the field at the first horizon
crossing, $\phi_i$, and at the end of inflation, $\phi_f$, have
been computed numerically. As usual, 60 $e$-foldings have been
assumed to occur between the two events. No further
simplifications or assumptions have been used. The Table
\ref{tab:LPIsets} presents 4 sets that respect the Planck data and
demonstrates that it is possible to achieve a tensor spectral
index as high as, $n_{\mathcal{T}}=0.9$, in three cases, namely
sets 5-7. Two regions are distinguished, $0\lesssim
n_{\mathcal{T}}\lesssim 0.45$ and $0.7\lesssim
n_{\mathcal{T}}\lesssim 1$, but the analysis holds the same, as
the behavior of the indices is similar.
\begin{table}
\centering
\begin{tabular}{|c|c|c|c|c|c|c|c|}
\hline
 & $\boldsymbol{p}$ & $\boldsymbol{q}$ & $\boldsymbol{\phi_0 }(10^{18}GeV)$ & $\boldsymbol{\beta}$ & $\boldsymbol{n_{\mathcal{S}}}$ & $\boldsymbol{n_{\mathcal{T}}}$ & $\boldsymbol{r}$ \\ \hline
\textbf{set 1} & 0.1  & 6 & $5\times10^9$ & 0.6  & 0.964 & 0.287 &
$1.06\times10^{-66}$   \\ \hline \textbf{set 2} & 0.4  & 6 &
$6\times10^6$ & 0.8  & 0.965  & 0.468 & $6.34\times10^{-20}$  \\
\hline \textbf{set 3} & 1  & 6   &    $10^6$    & 1    & 0.964   &
0.384 & 0.001  \\ \hline \textbf{set 4} & 1  & 10  & $3\times10^4$
& 0.95  & 0.965 & 0.447 & $3.9\times10^{-13}$   \\ \hline
\textbf{set 5} & 1  & 6 & $2\times10^6$  & 0.87  & 0.963 & 0.71 &
$9.2\times10^{-6}$   \\ \hline \textbf{set 6}  & 2  & 10 & $10^5$
& 0.9  & 0.964 & 0.91 & $9.2\times10^{-6}$
\\ \hline \textbf{set 7} & 3  & 8 & $1.1\times10^6$  & 0.93  &
0.963 & 0.93 & 0.057\\\hline
\end{tabular}
\caption{Table of viable cases for the LPI scenario.}
\label{tab:LPIsets}
\end{table}
In most cases, $q=6$ was a crucial choice, as it made it possible
for the spectral indices to acquire meaningful values, but $q=8$
and $q=10$ also proved to work. The region $0<p\leq1$ was most
suitable in order to keep a low $r$ and a sensible tensor
blue-tilt, $0.2\lesssim n_{\mathcal{T}}\lesssim 0.45$, while $p=2$
and $p=3$ where most suitable. For lower $n_{\mathcal{T}}$ and
while $0<p<1$, the tensor-to-scalar ratio is significantly lower
(respecting the Planck data upper limit, $r<0.064$ and also the
limit $r<0.036$ indicated by the BICEP/Keck experiment
\cite{BICEP:2021xfz}, with the only case being incompatible being
set 7 in Table \ref{tab:LPIsets}), but it exceeds this constraint
at higher values of $p$, as it is qualitatively shown in Figs.
(\ref{fig:LPIextraplots1}) and (\ref{fig:LPIextraplots2}).
\begin{figure}
    \centering
    \includegraphics[width=0.48\textwidth]{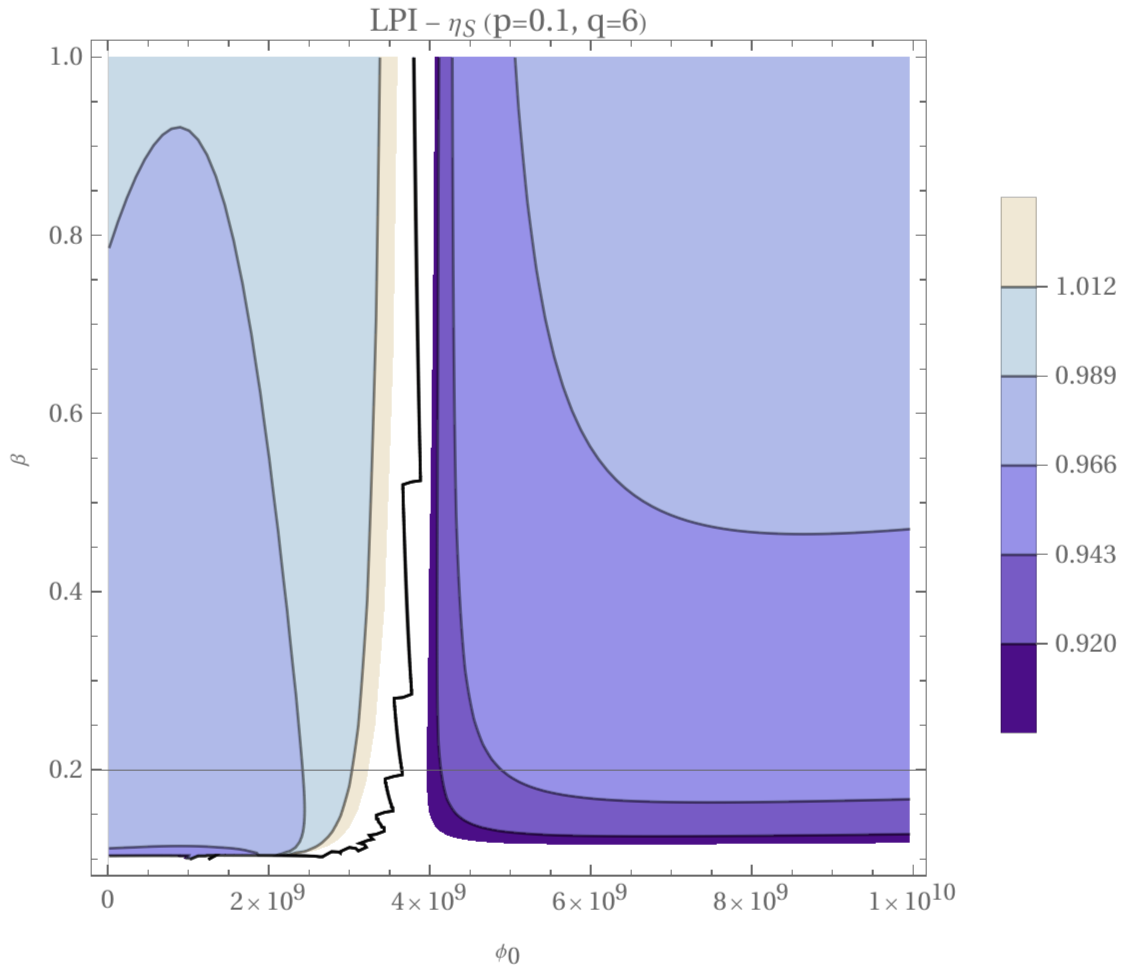}
    \hfill
    \includegraphics[width=0.48\textwidth]{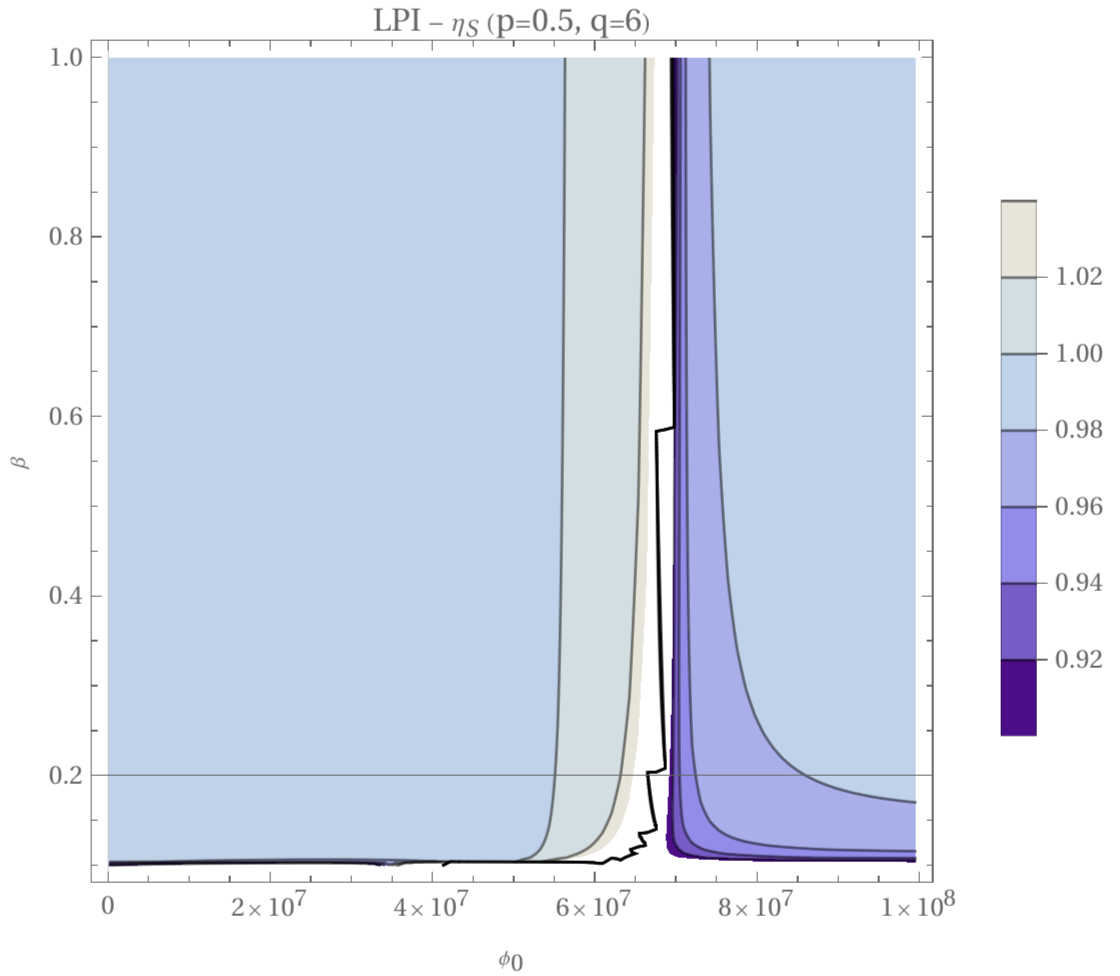}
    \vspace*{1cm}
    \includegraphics[width=0.48\textwidth]{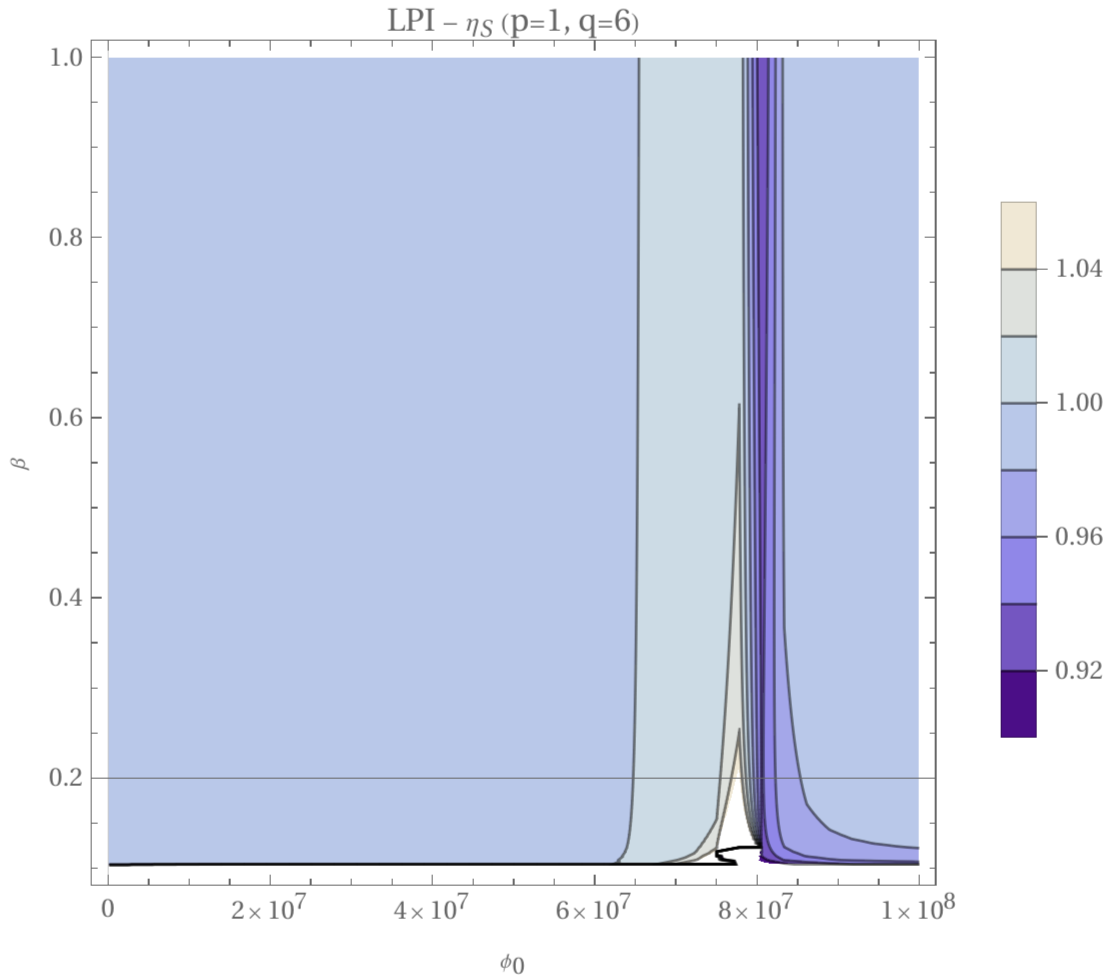}
    \caption{
        The figure displays the comparison between the $n_{\mathcal{S}}$ for three different cases:
        (a) Contour plot of scalar spectral index for $p=0.1$,
        (b) Contour plot of scalar spectral index for $p=0.5$, and
        (c) Contour plot of scalar spectral index for $p=1$.
        In all cases, just after the transition region where the function does not behave smoothly, the parameter $\phi_0$ is on its applicable scale.
    }
    \label{fig:LPInsplots}
\end{figure}

\begin{figure}
    \centering
    \includegraphics[width=0.45\textwidth]{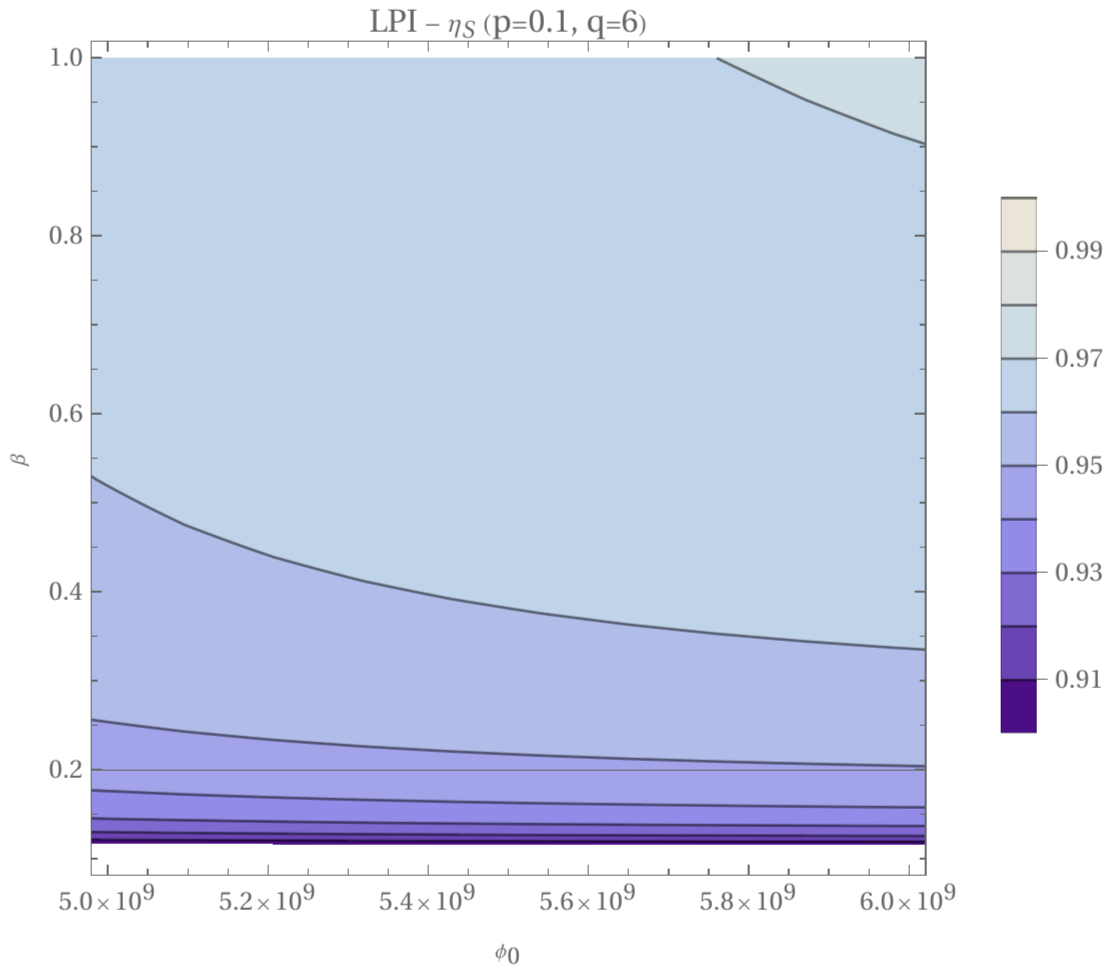}
    \hfill
    \includegraphics[width=0.45\textwidth]{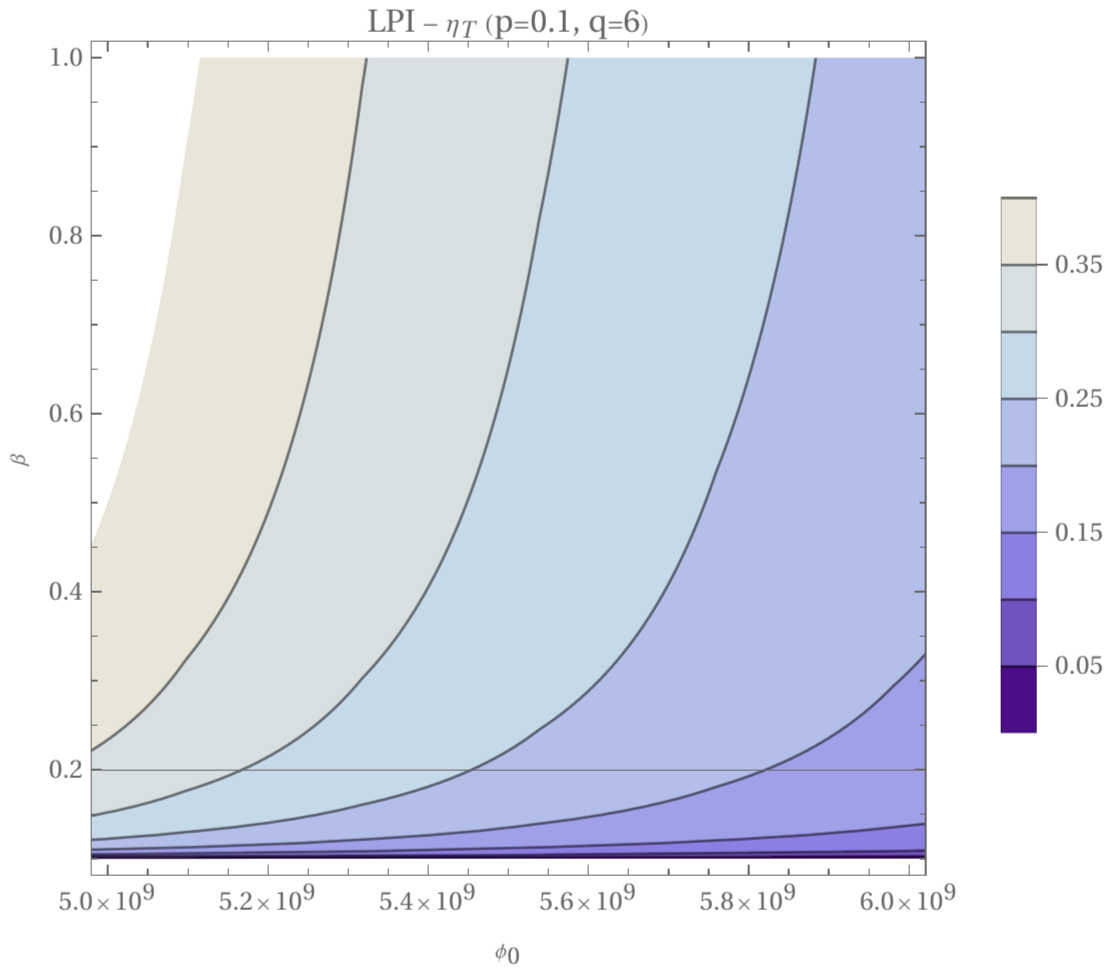}
    \caption{
This figure displays the contour plots of (a) Scalar spectral
index $n_{\mathcal{S}}$ and (b) Tensor spectral index
$n_{\mathcal{T}}$ with $\phi_0$ restricted to the region of
interest, for the case $p=0.1$,. The remaining plots are similar.
    }
    \label{fig:LPIclarity}
\end{figure}

\begin{figure}
    \centering
    \includegraphics[width=0.45\textwidth]{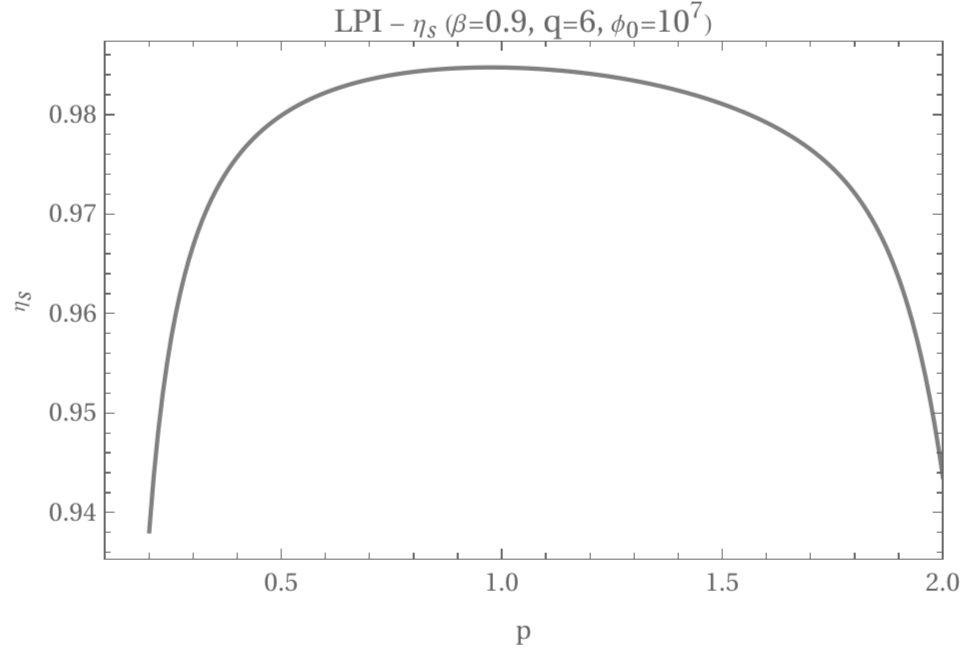}
    \hfill
    \includegraphics[width=0.45\textwidth]{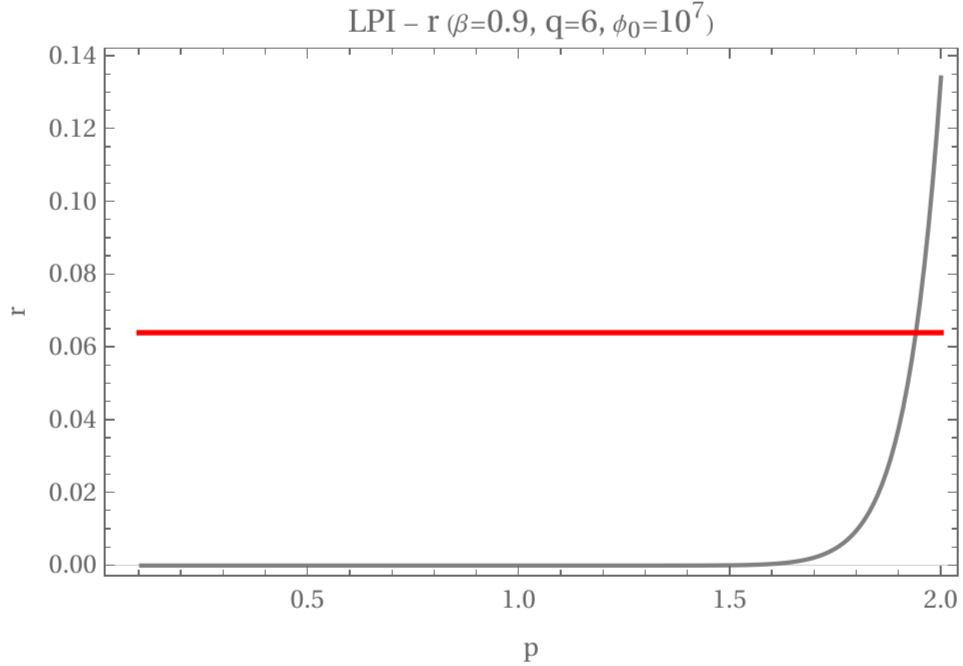}
    \caption{
        The figure shows two plots for $q=6$, $\phi_0=10^7$ and $\beta=0.9$:
        (a) Plot of the scalar spectral index $n_{\mathcal{S}}$ as it changes with $p$, and
        (b) Plot of the tensor-scalar-ratio $r$ as it changes with $p$, with the red line denoting the Planck constraint $r < 0.064$.
    }
    \label{fig:LPIextraplots}
\end{figure}

\begin{figure}[h!]
\centering
\includegraphics[width=20pc]{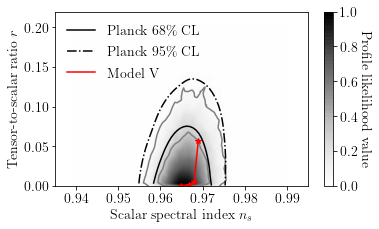}
\caption{The phenomenology of model V confronted with the Planck
2018 likelihood curves for various optimal values of the free
parameters.} \label{likelihoodmodel5}
\end{figure}

The scale at which $\phi_0$ becomes viable for our theory, at each
choice of $p$ and $q$, is clearly visible in the contour plots
(Fig. (\ref{fig:LPInsplots})).  The behavior is also similar for
the sets 5,6 and 7 in the same range $[10^5-10^7]$. The contours
seem to have weak dependence on $\beta$, but this is an illusion
caused by the range scale of the parameter $\phi_0$. This is the
reason an extra exemplary contour has been added (Sub-figures
\ref{fig:LPIclarity} magnified in a short interval of $\phi_0$
that is in accordance to the 2018 Planck constraints.
Furthermore, one can see how $n_{\mathcal{S}}$ and $r$ evolve with
the power parameter $p$, which strongly suggests that $p<2$
(Figures \ref{fig:LPIextraplots}(a) and
\ref{fig:LPIextraplots}(b)) In general, $p\geq2$ requires
$n_{\mathcal{T}}>0.5$. This model needs some care while picking
the parameter $q$, as the indices are very sensitive to small
changes in the power of the logarithm (especially comparing to the
other parameters) and only discrete choices present sensible
results.
Lastly, in order to confirm that the analysis lies within the
slow-roll regime, the initial hypotheses has to be validated,
which, actually, is the case here, as depicted in the Table
\ref{tab:LPIvalidation}. Unfortunately, in some cases $\epsilon_2$
is close to $0.9$. Also, the approximation $r=16\epsilon_1$ holds
up to many orders. The fact that the slow-roll indices at first
horizon crossing are taking such large values, makes this model
phenomenologically unappealing and we need to mention this, since
this model yields too large blue tilt for the tensor spectral
index. Thus we cannot consider this seriously since it is a
marginally correct model.
\begin{table}
\centering
\begin{tabular}{|c|c|c|c|c|c|c|}
\hline
 & $\boldsymbol{\epsilon_1}$ & $\boldsymbol{\epsilon_2}$ & $\boldsymbol{\epsilon_4}$ & $\boldsymbol{\epsilon_6}$ & $\boldsymbol{\kappa \frac{\xi'}{\xi''}}$ & $\boldsymbol{\frac{4\kappa^{4}\xi'^{2}V}{3\beta^{2}\xi''}}$ \\ \hline
\textbf{set 1} & $\mathcal{O}(10^{-68})$ & $\mathcal{O}(10^{-1})$ & $\mathcal{O}(10^{-1})$ & $\mathcal{O}(10^{-1})$ & $\mathcal{O}(10^{-34})$ & $\mathcal{O}(10^{}-1)$ \\ \hline
\textbf{set 2} &  $\mathcal{O}(10^{-21})$ & $\mathcal{O}(10^{-1})$& $\mathcal{O}(10^{-1})$ & $\mathcal{O}(10^{-1})$ &$\mathcal{O}(10^{-11})$ & $\mathcal{O}(10^{-1})$ \\ \hline
\textbf{set 3} &  $\mathcal{O}(10^{-5})$ & $\mathcal{O}(10^{-1})$ & $\mathcal{O}(10^{-1})$  & $\mathcal{O}(10^{-1})$&  $\mathcal{O}(10^{-2})$  & $\mathcal{O}(10^{-1})$\\ \hline
\textbf{set 4} & $\mathcal{O}(10^{-14})$& $\mathcal{O}(10^{-1})$ &$\mathcal{O}(10^{-1})$& $\mathcal{O}(10^{-1})$ & $\mathcal{O}(10^{-7})$ & $\mathcal{O}(10^{-1})$ \\ \hline
\textbf{set 5} & $\mathcal{O}(10^{-7})$& $\mathcal{O}(10^{-1})$ &$\mathcal{O}(10^{-1})$& $\mathcal{O}(10^{-1})$ & $\mathcal{O}(10^{-3})$ & $\mathcal{O}(10^{-1})$ \\ \hline
\textbf{set 6} & $\mathcal{O}(10^{-9})$& $\mathcal{O}(10^{-1})$ &$\mathcal{O}(10^{-1})$& $\mathcal{O}(10^{-1})$ & $\mathcal{O}(10^{-5})$ & $\mathcal{O}(10^{-1})$ \\ \hline
\textbf{set 7} & $\mathcal{O}(10^{-3})$& $\mathcal{O}(10^{-1})$ &$\mathcal{O}(10^{-1})$& $\mathcal{O}(10^{-1})$ & $\mathcal{O}(10^{-2})$ & $\mathcal{O}(10^{-1})$ \\ \hline
\end{tabular}
\caption{Validation of the slow-roll approximations for the LPI
inflation scenario.} \label{tab:LPIvalidation}
\end{table}
In this section we considered several rescaled models of EGB
gravity and one good question is when do the models yield a blue
tilt in the tensor spectral index. This is of particular
importance since a blue tilt might lead to detectable primordial
gravitational waves as we will show in the next section. However,
we need to note that the blue tilt in the tensor spectral index
must be strong enough and also the model must be viable regarding
the Planck data. From the models we considered, only model I and
model IV yield a strong blue tilt. In general, the blue tilt in
the tensor spectral index is a model dependent feature, but there
is a general rule on how such a blue tilt might occur. The method
for examining this was developed in Ref. \cite{Oikonomou:2024aww}
and at this point we shall discuss the results in brief.
Particularly, the tensor spectral index takes the form,
$n_{\mathcal{T}}\simeq 2\left(-1+\frac{1}{\lambda(\phi)}
\right)\epsilon_1$ at leading order, where $\lambda (\phi)$ is
defined in Eq. (\ref{eq:25}) and recall that it also contains the
rescaling parameter $\beta$. As it was shown in
\cite{Oikonomou:2024aww}, the blue tilted tensor spectral index is
obtained whenever $\xi''(\phi_*)V(\phi_*)>0$ at first horizon
crossing. So model building may be used in such a way so that the
constraint $\xi''(\phi_*)V(\phi_*)>0$ is satisfied at first
horizon crossing, and simultaneously the model must be compatible
with the Planck and BICEP/Keck data. Furthermore, in Fig.
\ref{likelihoodmodel5} we confront model II with the Planck 2018
likelihood curves for various suitably chosen values of the free
parameters. As it can be seen, the model is nicely fitted in the
Planck likelihood curves and for some values of the free
parameters, the model is phenomenologically fitted in the sweet
spot of the Planck data.

\section{Primordial Gravitational Wave Energy Spectrum for Rescaled EGB Theories}

In the previous sections we developed the theoretical framework of
rescaled EGB gravity, and we studied several models of interest
which were viable phenomenologically. More importantly, some of
these yielded a blue-tilted tensor spectrum, while being
simultaneously compatible with the Planck constraints on the
scalar spectral index of primordial perturbations and the
tensor-to-scalar ratio. In this section we shall investigate
whether some of these models can yield a detectable signal of
primordial gravitational waves, so we shall calculate the energy
spectrum of the primordial gravitational waves for the models of
interest. In the literature there is a large stream of articles
studying the primordial gravitational waves, see for example In
the literature there exist many works which consider theoretical
predictions on primordial gravitational waves, for a mainstream of
articles see for example Refs.
\cite{Kamionkowski:2015yta,Turner:1993vb,Boyle:2005se,Zhang:2005nw,Caprini:2018mtu,Clarke:2020bil,Smith:2005mm,Giovannini:2008tm,Liu:2015psa,Vagnozzi:2020gtf,Giovannini:2023itq,Giovannini:2022eue,Giovannini:2022vha,Giovannini:2020wrx,Giovannini:2019oii,Giovannini:2019ioo,Giovannini:2014vya,Giovannini:2009kg,Kamionkowski:1993fg,Giare:2020vss,Zhao:2006mm,Lasky:2015lej,
Cai:2021uup,Odintsov:2021kup,Lin:2021vwc,Zhang:2021vak,Visinelli:2017bny,Pritchard:2004qp,Khoze:2022nyt,Casalino:2018tcd,Oikonomou:2022xoq,Casalino:2018wnc,ElBourakadi:2022anr,Sturani:2021ucg,Vagnozzi:2022qmc,Arapoglu:2022vbf,Giare:2022wxq,Oikonomou:2021kql,Gerbino:2016sgw,Breitbach:2018ddu,Pi:2019ihn,Khlopov:2023mpo,Odintsov:2022cbm,Benetti:2021uea,Vagnozzi:2020gtf}.
The models which are interesting phenomenologically, are the power
law model of Eq. (\ref{eq:32}) in which case, the tensor spectral
index takes the value $n_{\mathcal{T}}=0.56$ and also $r=0.02$ for
the set 2 appearing in Table \ref{tab:1}. Also, another model we
shall consider is the logarithmic model of inflation
(\ref{logpotential}) and specifically the set 6 from Table
\ref{tab:LPIsets}, which yields $n_{\mathcal{T}}=0.91$ and
$r=9.2\times10^{-6}$, however this model has to be treated with
caution since the approximations marginally hold true.
\begin{figure}[h!]
\centering
\includegraphics[width=40pc]{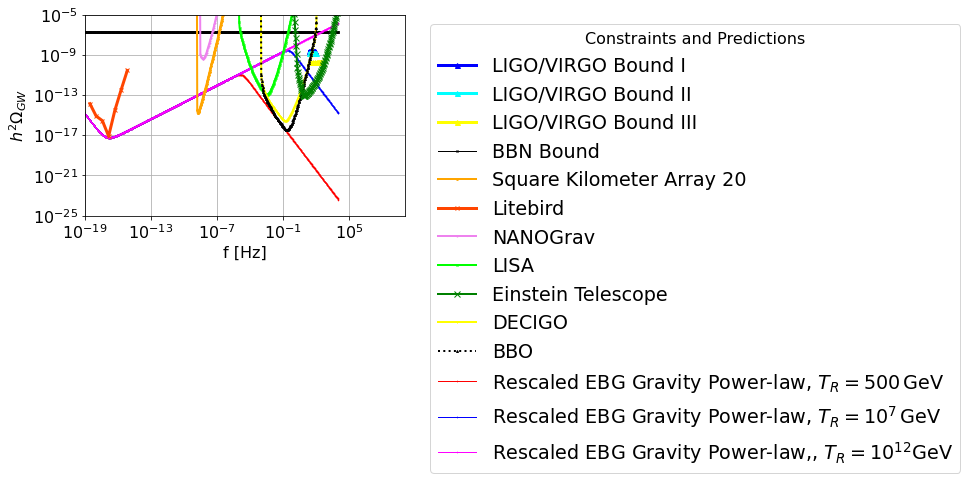}
\caption{The $h^2$-scaled gravitational wave energy spectrum for
the rescaled EGB power-law model of gravity (\ref{eq:32}) for
three distinct reheating temperatures, $T_R=5\times 10^{2}$GeV,
$T_R=10^{7}$GeV and $T_R=10^{12}$GeV.} \label{plotfinalfrpure1}
\end{figure}
The energy spectrum today of an inflationary theory is,
\begin{equation}
    \Omega_{\rm gw}(f)= \frac{k^2}{12H_0^2}\Delta_h^2(k),
    \label{GWspec}
\end{equation}
with $\Delta_h^2(k)$ defined as,
\begin{equation}\label{mainfunctionforgravityenergyspectrum}
    \Delta_h^2(k)=\Delta_h^{({\rm p})}(k)^{2}
    \left ( \frac{\Omega_m}{\Omega_\Lambda} \right )^2
    \left ( \frac{g_*(T_{\rm in})}{g_{*0}} \right )
    \left ( \frac{g_{*s0}}{g_{*s}(T_{\rm in})} \right )^{4/3} \nonumber  \left (\overline{ \frac{3j_1(k\tau_0)}{k\tau_0} } \right )^2
    T_1^2\left ( x_{\rm eq} \right )
    T_2^2\left ( x_R \right ),
\end{equation}
while $\Delta_h^{({\rm p})}(k)^{2}$ stands for the inflationary
tensor power spectrum, which is defined as,
\begin{equation}\label{primordialtensorpowerspectrum}
\Delta_h^{({\rm
p})}(k)^{2}=\mathcal{A}_T(k_{ref})\left(\frac{k}{k_{ref}}
\right)^{n_{T}}\, .
\end{equation}
We evaluate the inflationary tensor power spectrum  at the CMB
pivot scale $k_{ref}=0.002$$\,$Mpc$^{-1}$ and $n_{T}$ denotes the
spectral index of the tensor perturbations, while
$\mathcal{A}_T(k_{ref})$ stands for the amplitude of the tensor
perturbations, which is,
\begin{equation}\label{amplitudeoftensorperturbations}
\mathcal{A}_T(k_{ref})=r\mathcal{P}_{\zeta}(k_{ref})\, .
\end{equation}
Also $r$ is the tensor-to-scalar ratio and
$\mathcal{P}_{\zeta}(k_{ref})$ denotes the amplitude of the
primordial scalar perturbations. Therefore, we have,
\begin{equation}\label{primordialtensorspectrum}
\Delta_h^{({\rm
p})}(k)^{2}=r\mathcal{P}_{\zeta}(k_{ref})\left(\frac{k}{k_{ref}}
\right)^{n_{\mathcal{T}}}\, ,
\end{equation}
hence the energy spectrum of the primordial gravitational waves
which we shall calculate for the rescaled EGB theories reads,
\begin{align}
\label{GWspecfR}
    &\Omega_{\rm gw}(f)=\frac{k^2}{12H_0^2}r\mathcal{P}_{\zeta}(k_{ref})\left(\frac{k}{k_{ref}}
\right)^{n_{\mathcal{T}}} \left ( \frac{\Omega_m}{\Omega_\Lambda}
\right )^2
    \left ( \frac{g_*(T_{\rm in})}{g_{*0}} \right )
    \left ( \frac{g_{*s0}}{g_{*s}(T_{\rm in})} \right )^{4/3} \nonumber  \left (\overline{ \frac{3j_1(k\tau_0)}{k\tau_0} } \right )^2
    T_1^2\left ( x_{\rm eq} \right )
    T_2^2\left ( x_R \right )\, ,
\end{align}
where $T_{\rm in}$ denotes the horizon reentry temperature,
\begin{equation}
    T_{\rm in}\simeq 5.8\times 10^6~{\rm GeV}
    \left ( \frac{g_{*s}(T_{\rm in})}{106.75} \right )^{-1/6}
    \left ( \frac{k}{10^{14}~{\rm Mpc^{-1}}} \right )\, ,
\end{equation}
and also the transfer function $T_1(x_{\rm eq})$ is defined to be
equal to,
\begin{equation}
    T_1^2(x_{\rm eq})=
    \left [1+1.57x_{\rm eq} + 3.42x_{\rm eq}^2 \right ], \label{T1}
\end{equation}
where $x_{\rm eq}=k/k_{\rm eq}$ and $k_{\rm eq}\equiv a(t_{\rm
eq})H(t_{\rm eq}) = 7.1\times 10^{-2} \Omega_m h^2$ Mpc$^{-1}$,
and in addition the transfer function $T_2(x_R)$ is equal to,
\begin{equation}\label{transfer2}
 T_2^2\left ( x_R \right )=\left(1-0.22x^{1.5}+0.65x^2
 \right)^{-1}\, ,
\end{equation}
with $x_R=\frac{k}{k_R}$, and in addition the wavenumber at the
epoch of maximum reheating is,
\begin{equation}
    k_R\simeq 1.7\times 10^{13}~{\rm Mpc^{-1}}
    \left ( \frac{g_{*s}(T_R)}{106.75} \right )^{1/6}
    \left ( \frac{T_R}{10^6~{\rm GeV}} \right )\, ,  \label{k_R}
\end{equation}
where $T_R$ denotes the reheating temperature, which is very
important in our consideration, and we shall discuss its values
later on in this section. Also, $g_*(T_{\mathrm{in}}(k))$ is
defined as,
\begin{align}\label{gstartin}
& g_*(T_{\mathrm{in}}(k))=g_{*0}\left(\frac{A+\tanh \left[-2.5
\log_{10}\left(\frac{k/2\pi}{2.5\times 10^{-12}\mathrm{Hz}}
\right) \right]}{A+1} \right) \left(\frac{B+\tanh \left[-2
\log_{10}\left(\frac{k/2\pi}{6\times 10^{-19}\mathrm{Hz}} \right)
\right]}{B+1} \right)\, ,
\end{align}
where $A$ and $B$ parameters are equal to,
\begin{equation}\label{alphacap}
A=\frac{-1-10.75/g_{*0}}{-1+10.75g_{*0}}\, ,
\end{equation}
\begin{equation}\label{betacap}
B=\frac{-1-g_{max}/10.75}{-1+g_{max}/10.75}\, ,
\end{equation}
with $g_{max}=106.75$ and $g_{*0}=3.36$. Furthermore,
$g_{*0}(T_{\mathrm{in}}(k))$ and can be evaluated by using Eqs.
(\ref{gstartin}), (\ref{alphacap}) and (\ref{betacap}), by
replacing $g_{*0}=3.36$ with $g_{*s}=3.91$.
\begin{figure}[h!]
\centering
\includegraphics[width=40pc]{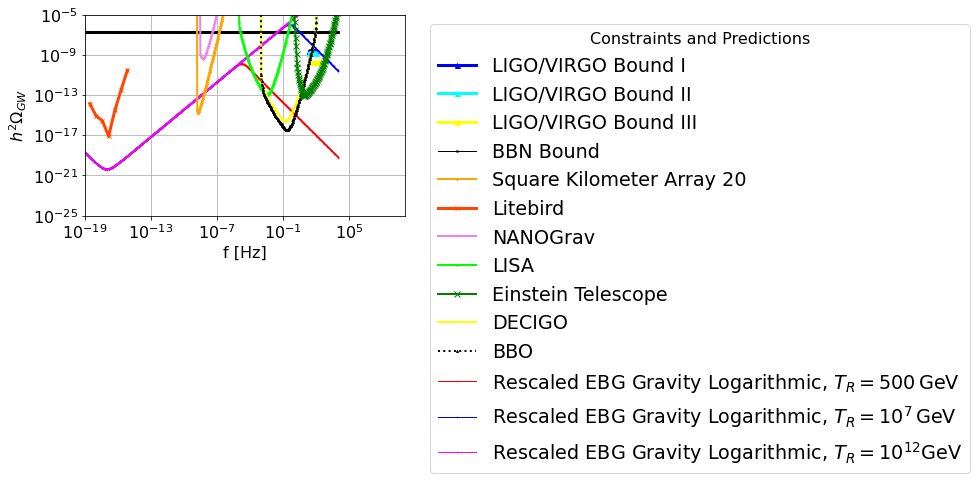}
\caption{The $h^2$-scaled gravitational wave energy spectrum for
the rescaled EGB logarithmic model of gravity (\ref{logpotential})
for three distinct reheating temperatures, $T_R=5\times
10^{2}$GeV, $T_R=10^{7}$GeV and $T_R=10^{12}$GeV.}
\label{plotfinalfrpure2}
\end{figure}
Now let us proceed by evaluating the $h^2$-scaled energy spectrum
of the primordial gravitational waves for the rescaled EGB models
we mentioned earlier, namely the  power law model of Eq.
(\ref{eq:32}) and the logarithmic model (\ref{logpotential}). We
shall consider three reheating temperatures, namely a high
reheating temperature $T_R=10^{12}$GeV, an intermediate one with
$T_R=10^{7}$GeV and a low reheating temperature $T_R=5\times
10^{2}$GeV. In Fig. \ref{plotfinalfrpure1} we present the
predicted $h^2$-scaled energy spectrum of the primordial
gravitational waves for the power-law model (\ref{eq:32}) for
$n_{\mathcal{T}}=0.56$ and also $r=0.02$, while in Fig.
\ref{plotfinalfrpure2} we present the predicted $h^2$-scaled
energy spectrum of the primordial gravitational waves for the
logarithmic model (\ref{eq:32}) for $n_{\mathcal{T}}=0.91$ and
$r=9.2\times10^{-6}$. As it can be seen, the power-law model leads
to a detectable signal which can be detected by the SKA, LISA,
BBO, DECIGO and the Einstein Telescope, for intermediate reheating
temperatures. The high reheating temperature scenario is excluded
by the LIGO constraints, while the low-reheating temperature
scenario can be detectable only by the SKA experiment. Regarding
the logarithmic model, only the low-reheating temperature scenario
may be detectable, while the high and intermediate reheating
temperature scenarios are excluded by the current LIGO
constraints. In this scenario, the low-reheating case can be
detectable by LISA, SKA, BBO and DECIGO. Thus the phenomenology of
rescaled EGB gravity is quite promising and can lead to
particularly interesting results which may be verified by future
gravitational wave experiments.

\section{Concluding Remarks}

In this work, we employed a self-consistent and simplified
rescaled EGB inflationary theoretical framework, and we analyzed
several models of theoretical interest. We developed the formalism
of the rescaled EGB theories of gravity, and we demonstrated how
the rescaled theory can directly affect the primordial scalar and
tensor perturbations. Accordingly, we showed how the slow-roll
indices will be affected and eventually how the inflationary
observational indices are changed. We applied the formalism for
specifically chosen models and we showed how these models can
provide a viable inflationary phenomenology. Specifically, the
most important models that we analyzed are the natural inflation
potential extended in an EGB theory, and a power-law scalar
potential. As we demonstrated, the rescaled versions of these
models can provide a viable inflationary theory, compatible with
the latest Planck data and more importantly these models can lead
to a significantly blue-tilted tensor spectral index, a feature
that is phenomenologically appealing since blue-tilted
inflationary theories can lead to a detectable signal of
primordial gravitational waves. We calculated the energy spectrum
of the primordial gravitational waves for the models of interest
and we showed that, depending on the reheating temperature, these
models can lead to detectable stochastic signals in the future
gravitational wave detectors, and specifically some signals can be
detected by LISA, SKA, BBO, DECIGO and the Einstein Telescope.
What we did not consider is the dark energy implications of this
rescaled EGB theories. Basically, the rescaled EGB theories
contain a non-trivial $F(R)$ gravity part, which primordially
contributes only the rescaling of the Einstein-Hilbert term, but
at late times it can become significant since the curvature is not
large anymore. Thus, phenomena like the ones described in Ref.
\cite{Oikonomou:2020oex} might occur, in the presence of a
non-trivial Gauss-Bonnet coupling in the inflaton. This is of
particular importance since the $F(R)$ gravity part may
significantly affect the late-time evolution of the EGB theory. In
this paper we considered a non-trivial part for the $F(R)$ gravity
of exponential form $e^{-\Lambda/R}$ which can easily dominate the
late-time evolution. On the contrary, primordially this
exponential term contributes nothing more than a rescaling in the
Einstein-Hilbert term, which affects the EGB phenomenology at the
first horizon crossing. The rescaling of the Einstein-Hilbert term
becomes negligible near the end of inflation, after the Universe
has expanded significantly and the Ricci scalar basically takes
small values. Hence the rescaling only occurs near the first
horizon crossing and does not occur during the subsequent
evolution eras of our Universe, from the end of inflation, and
after, including the reheating and radiation era and thereafter.
Such rescaling theories can be considered in the context of other
modified gravities, like for example $f(R,T)$ or $F(R,G)$, in the
latter theories however the problems of ghosts must be considered.
In the present context ghosts are not a problem because the theory
is an $F(R)$ gravity extension of the standard EGB theory, which
is known to be ghost free in the linear perturbations of the
metric context. In addition, one may consider such rescaled $F(R)$
theories in strong gravity regimes, such as near black holes or in
neutron stars. We have not considered these effects here, but it
certainly would be interesting to address such problems in a study
focused on compact objects with extreme gravity.

Before closing this article, we need to discuss an important issue
related with the duration of the reheating era. As it was
demonstrated in Ref. \cite{Pi:2024kpw}, the equation of state
(EoS) of the Universe at the post-inflationary era may affect the
energy spectrum of the primordial gravitational waves in a direct
way. As it is discussed in \cite{Pi:2024kpw}, the transition from
inflation to an intermediate EoS Universe, before the Universe
enters its radiation domination EoS $w=1/3$, can have an imprint
in the form of a peak on the energy spectrum of the primordial
gravitational waves. This important aspect was not taken into
account in this work, but we aim to discuss this issue soon in the
context of both $F(R)$ gravity and EGB gravity. Note that the
transition from an inflationary de Sitter epoch to the radiation
domination era in the context of simple EGB theories, was studied
in Ref. \cite{Oikonomou:2024jqv}, so these considerations must
also be taken into account when considering the duration and total
background EoS in the context of EGB theories.

\section*{Acknowledgments}

This research has been is funded by the Committee of Science of
the Ministry of Education and Science of the Republic of
Kazakhstan (Grant No. AP19674478).


\begin{thebibliography}{99}



\bibitem{inflation1}
 A.~D.~Linde,
 Lect.\ Notes Phys.\ {\bf 738} (2008) 1
 [arXiv:0705.0164 [hep-th]].

\bibitem{inflation2} D.~S.~Gorbunov and V.~A.~Rubakov,
``Introduction to the theory of the early universe: Cosmological
perturbations and inflationary theory,'' Hackensack, USA: World
Scientific (2011) 489 p;
%


\bibitem{inflation3}A.~Linde,
arXiv:1402.0526 [hep-th];


\bibitem{inflation4}D.~H.~Lyth and A.~Riotto,
Phys.\ Rept.\  {\bf 314} (1999) 1 [hep-ph/9807278].


\bibitem{SimonsObservatory:2019qwx}
M.~H.~Abitbol \textit{et al.} [Simons Observatory],
Bull. Am. Astron. Soc. \textbf{51} (2019), 147 [arXiv:1907.08284
[astro-ph.IM]].



\bibitem{CMB-S4:2016ple}
K.~N.~Abazajian \textit{et al.} [CMB-S4],
[arXiv:1610.02743 [astro-ph.CO]].









\bibitem{Hild:2010id}
S.~Hild, M.~Abernathy, F.~Acernese, P.~Amaro-Seoane, N.~Andersson,
K.~Arun, F.~Barone, B.~Barr, M.~Barsuglia and M.~Beker, \textit{et
al.}
Class. Quant. Grav. \textbf{28} (2011), 094013
doi:10.1088/0264-9381/28/9/094013 [arXiv:1012.0908 [gr-qc]].




\bibitem{Baker:2019nia}
J.~Baker, J.~Bellovary, P.~L.~Bender, E.~Berti, R.~Caldwell,
J.~Camp, J.~W.~Conklin, N.~Cornish, C.~Cutler and R.~DeRosa,
\textit{et al.}
[arXiv:1907.06482 [astro-ph.IM]].


\bibitem{Smith:2019wny}
T.~L.~Smith and R.~Caldwell,
Phys. Rev. D \textbf{100} (2019) no.10, 104055
doi:10.1103/PhysRevD.100.104055 [arXiv:1908.00546 [astro-ph.CO]].


\bibitem{Crowder:2005nr}
J.~Crowder and N.~J.~Cornish,
Phys. Rev. D \textbf{72} (2005), 083005
doi:10.1103/PhysRevD.72.083005 [arXiv:gr-qc/0506015 [gr-qc]].


\bibitem{Smith:2016jqs}
T.~L.~Smith and R.~Caldwell,
Phys. Rev. D \textbf{95} (2017) no.4, 044036
doi:10.1103/PhysRevD.95.044036 [arXiv:1609.05901 [gr-qc]].



\bibitem{Seto:2001qf}
N.~Seto, S.~Kawamura and T.~Nakamura,
Phys. Rev. Lett. \textbf{87} (2001), 221103
doi:10.1103/PhysRevLett.87.221103 [arXiv:astro-ph/0108011
[astro-ph]].


\bibitem{Kawamura:2020pcg}
S.~Kawamura, M.~Ando, N.~Seto, S.~Sato, M.~Musha, I.~Kawano,
J.~Yokoyama, T.~Tanaka, K.~Ioka and T.~Akutsu, \textit{et al.}
[arXiv:2006.13545 [gr-qc]].



\bibitem{Bull:2018lat}
A.~Weltman, P.~Bull, S.~Camera, K.~Kelley, H.~Padmanabhan,
J.~Pritchard, A.~Raccanelli, S.~Riemer-S\o{}rensen, L.~Shao and
S.~Andrianomena, \textit{et al.}
Publ. Astron. Soc. Austral. \textbf{37} (2020), e002
doi:10.1017/pasa.2019.42 [arXiv:1810.02680 [astro-ph.CO]].




\bibitem{LISACosmologyWorkingGroup:2022jok}
P.~Auclair \textit{et al.} [LISA Cosmology Working Group],
[arXiv:2204.05434 [astro-ph.CO]].


\bibitem{NANOGrav:2023gor}
G.~Agazie \textit{et al.} [NANOGrav],
Astrophys. J. Lett. \textbf{951} (2023) no.1, L8
doi:10.3847/2041-8213/acdac6 [arXiv:2306.16213 [astro-ph.HE]].



\bibitem{Vagnozzi:2023lwo}
S.~Vagnozzi,
JHEAp \textbf{39} (2023), 81-98 doi:10.1016/j.jheap.2023.07.001
[arXiv:2306.16912 [astro-ph.CO]].


\bibitem{Yi:2023mbm}
Z.~Yi, Q.~Gao, Y.~Gong, Y.~Wang and F.~Zhang,
Sci. China Phys. Mech. Astron. \textbf{66} (2023) no.12, 120404
doi:10.1007/s11433-023-2266-1 [arXiv:2307.02467 [gr-qc]].


\bibitem{Balaji:2023ehk}
S.~Balaji, G.~Dom\`enech and G.~Franciolini,
JCAP \textbf{10} (2023), 041 doi:10.1088/1475-7516/2023/10/041
[arXiv:2307.08552 [gr-qc]].



\bibitem{Oikonomou:2023qfz}
V.~K.~Oikonomou,
Phys. Rev. D \textbf{108} (2023) no.4, 043516
doi:10.1103/PhysRevD.108.043516 [arXiv:2306.17351 [astro-ph.CO]].



\bibitem{reviews1}
 S.~Nojiri, S.~D.~Odintsov and V.~K.~Oikonomou,
  Phys.\ Rept.\  {\bf 692} (2017) 1
  [arXiv:1705.11098 [gr-qc]].

\bibitem{reviews2}


 S. Capozziello, M. De Laurentis,
   Phys.\ Rept.\  {\bf 509}, 167 (2011);\\
 V.~Faraoni and S.~Capozziello,
  Fundam.\ Theor.\ Phys.\  {\bf 170} (2010).



\bibitem{reviews3}
S. Nojiri, S.D. Odintsov,
  eConf {\bf C0602061}, 06 (2006)
  [Int.\ J.\ Geom.\ Meth.\ Mod.\ Phys.\  {\bf 4}, 115 (2007)].


   \bibitem{reviews4}

S. Nojiri, S.D. Odintsov,
   Phys.\ Rept.\  {\bf 505}, 59 (2011);






\bibitem{Hwang:2005hb}
  J.~c.~Hwang and H.~Noh,
  Phys.\ Rev.\ D {\bf 71} (2005) 063536
  doi:10.1103/PhysRevD.71.063536
  [gr-qc/0412126].


\bibitem{Nojiri:2006je}
  S.~Nojiri, S.~D.~Odintsov and M.~Sami,
  Phys.\ Rev.\ D {\bf 74} (2006) 046004
  doi:10.1103/PhysRevD.74.046004
  [hep-th/0605039].




\bibitem{Cognola:2006sp}
  G.~Cognola, E.~Elizalde, S.~Nojiri, S.~Odintsov and S.~Zerbini,
  Phys.\ Rev.\ D {\bf 75} (2007) 086002
  doi:10.1103/PhysRevD.75.086002
  [hep-th/0611198].



\bibitem{Nojiri:2005vv}
  S.~Nojiri, S.~D.~Odintsov and M.~Sasaki,
  Phys.\ Rev.\ D {\bf 71} (2005) 123509
  doi:10.1103/PhysRevD.71.123509
  [hep-th/0504052].


\bibitem{Nojiri:2005jg}
  S.~Nojiri and S.~D.~Odintsov,
  Phys.\ Lett.\ B {\bf 631} (2005) 1
  doi:10.1016/j.physletb.2005.10.010
  [hep-th/0508049].







\bibitem{Satoh:2007gn}
  M.~Satoh, S.~Kanno and J.~Soda,
  Phys.\ Rev.\ D {\bf 77} (2008) 023526
  doi:10.1103/PhysRevD.77.023526
  [arXiv:0706.3585 [astro-ph]].



\bibitem{Bamba:2014zoa}
  K.~Bamba, A.~N.~Makarenko, A.~N.~Myagky and S.~D.~Odintsov,
  JCAP {\bf 1504} (2015) 001
  doi:10.1088/1475-7516/2015/04/001
  [arXiv:1411.3852 [hep-th]].


\bibitem{Yi:2018gse}
  Z.~Yi, Y.~Gong and M.~Sabir,
  Phys.\ Rev.\ D {\bf 98} (2018) no.8,  083521
  doi:10.1103/PhysRevD.98.083521
  [arXiv:1804.09116 [gr-qc]].


\bibitem{Guo:2009uk}
  Z.~K.~Guo and D.~J.~Schwarz,
  Phys.\ Rev.\ D {\bf 80} (2009) 063523
  doi:10.1103/PhysRevD.80.063523
  [arXiv:0907.0427 [hep-th]].


\bibitem{Guo:2010jr}
  Z.~K.~Guo and D.~J.~Schwarz,
  Phys.\ Rev.\ D {\bf 81} (2010) 123520
  doi:10.1103/PhysRevD.81.123520
  [arXiv:1001.1897 [hep-th]].


\bibitem{Jiang:2013gza}
  P.~X.~Jiang, J.~W.~Hu and Z.~K.~Guo,
  Phys.\ Rev.\ D {\bf 88} (2013) 123508
  doi:10.1103/PhysRevD.88.123508
  [arXiv:1310.5579 [hep-th]].




\bibitem{vandeBruck:2017voa}
  C.~van de Bruck, K.~Dimopoulos, C.~Longden and C.~Owen,
  arXiv:1707.06839 [astro-ph.CO].



\bibitem{Pozdeeva:2020apf}
E.~O.~Pozdeeva, M.~R.~Gangopadhyay, M.~Sami, A.~V.~Toporensky and
S.~Y.~Vernov,
Phys. Rev. D \textbf{102} (2020) no.4, 043525
doi:10.1103/PhysRevD.102.043525 [arXiv:2006.08027 [gr-qc]].



\bibitem{Vernov:2021hxo}
S.~Vernov and E.~Pozdeeva,
Universe \textbf{7} (2021) no.5, 149 doi:10.3390/universe7050149
[arXiv:2104.11111 [gr-qc]].


\bibitem{Pozdeeva:2021iwc}
E.~O.~Pozdeeva and S.~Y.~Vernov,
Eur. Phys. J. C \textbf{81} (2021) no.7, 633
doi:10.1140/epjc/s10052-021-09435-8 [arXiv:2104.04995 [gr-qc]].


\bibitem{Fomin:2020hfh}
I.~Fomin,
Eur. Phys. J. C \textbf{80} (2020) no.12, 1145
doi:10.1140/epjc/s10052-020-08718-w [arXiv:2004.08065 [gr-qc]].

\bibitem{DeLaurentis:2015fea}
  M.~De Laurentis, M.~Paolella and S.~Capozziello,
  Phys.\ Rev.\ D {\bf 91} (2015) no.8,  083531
  doi:10.1103/PhysRevD.91.083531
  [arXiv:1503.04659 [gr-qc]].


\bibitem{Chervon:2019sey}
   Scalar Field Cosmology, S.~Chervon, I.~Fomin, V.~Yurov and
   A.~Yurov, World Scientific 2019, \\  doi:10.1142/11405



\bibitem{Nozari:2017rta}
  K.~Nozari and N.~Rashidi,
  Phys.\ Rev.\ D {\bf 95} (2017) no.12,  123518
  doi:10.1103/PhysRevD.95.123518
  [arXiv:1705.02617 [astro-ph.CO]].




\bibitem{Odintsov:2018zhw}
  S.~D.~Odintsov and V.~K.~Oikonomou,
  Phys.\ Rev.\ D {\bf 98} (2018) no.4,  044039
  doi:10.1103/PhysRevD.98.044039
  [arXiv:1808.05045 [gr-qc]].


  \bibitem{Kawai:1998ab}
  S.~Kawai, M.~a.~Sakagami and J.~Soda,
  Phys.\ Lett.\ B {\bf 437}, 284 (1998)
  doi:10.1016/S0370-2693(98)00925-3
  [gr-qc/9802033].


\bibitem{Yi:2018dhl}
  Z.~Yi and Y.~Gong,
  Universe {\bf 5} (2019) no.9,  200
  doi:10.3390/universe5090200
  [arXiv:1811.01625 [gr-qc]].


\bibitem{vandeBruck:2016xvt}
  C.~van de Bruck, K.~Dimopoulos and C.~Longden,
  Phys.\ Rev.\ D {\bf 94} (2016) no.2,  023506
  doi:10.1103/PhysRevD.94.023506
  [arXiv:1605.06350 [astro-ph.CO]].




\bibitem{Maeda:2011zn}
  K.~i.~Maeda, N.~Ohta and R.~Wakebe,
  Eur.\ Phys.\ J.\ C {\bf 72} (2012) 1949
  doi:10.1140/epjc/s10052-012-1949-6
  [arXiv:1111.3251 [hep-th]].


\bibitem{Ai:2020peo}
W.~Y.~Ai,
Commun. Theor. Phys. \textbf{72} (2020) no.9, 095402
doi:10.1088/1572-9494/aba242 [arXiv:2004.02858 [gr-qc]].





\bibitem{Easther:1996yd}
  R.~Easther and K.~i.~Maeda,
  Phys.\ Rev.\ D {\bf 54} (1996) 7252
  doi:10.1103/PhysRevD.54.7252
  [hep-th/9605173].



\bibitem{Codello:2015mba}
A.~Codello and R.~K.~Jain,
Class. Quant. Grav. \textbf{33} (2016) no.22, 225006
doi:10.1088/0264-9381/33/22/225006 [arXiv:1507.06308 [gr-qc]].







\bibitem{TheLIGOScientific:2017qsa}
B.~P.~Abbott \textit{et al.} [LIGO Scientific and Virgo],
Phys. Rev. Lett. \textbf{119} (2017) no.16, 161101
doi:10.1103/PhysRevLett.119.161101 [arXiv:1710.05832 [gr-qc]].




\bibitem{Monitor:2017mdv}
B.~P.~Abbott \textit{et al.} [LIGO Scientific, Virgo, Fermi-GBM
and INTEGRAL],
Astrophys. J. Lett. \textbf{848} (2017) no.2, L13
doi:10.3847/2041-8213/aa920c [arXiv:1710.05834 [astro-ph.HE]].


\bibitem{GBM:2017lvd}
  B.~P.~Abbott {\it et al.}
  ``Multi-messenger Observations of a Binary Neutron Star Merger,''
  Astrophys.\ J.\  {\bf 848} (2017) no.2,  L12
  doi:10.3847/2041-8213/aa91c9
  [arXiv:1710.05833 [astro-ph.HE]].


\bibitem{LIGOScientific:2019vic}
B.~P.~Abbott \textit{et al.} [LIGO Scientific and Virgo],
Phys. Rev. D \textbf{100} (2019) no.6, 061101
doi:10.1103/PhysRevD.100.061101 [arXiv:1903.02886 [gr-qc]].




\bibitem{Ezquiaga:2017ekz}
J.~M.~Ezquiaga and M.~Zumalac\'arregui,
Phys. Rev. Lett. \textbf{119} (2017) no.25, 251304
doi:10.1103/PhysRevLett.119.251304 [arXiv:1710.05901
[astro-ph.CO]].


\bibitem{Baker:2017hug}
T.~Baker, E.~Bellini, P.~G.~Ferreira, M.~Lagos, J.~Noller and
I.~Sawicki,
Phys. Rev. Lett. \textbf{119} (2017) no.25, 251301
doi:10.1103/PhysRevLett.119.251301 [arXiv:1710.06394
[astro-ph.CO]].


\bibitem{Creminelli:2017sry}
P.~Creminelli and F.~Vernizzi,
Phys. Rev. Lett. \textbf{119} (2017) no.25, 251302
doi:10.1103/PhysRevLett.119.251302 [arXiv:1710.05877
[astro-ph.CO]].


\bibitem{Sakstein:2017xjx}
J.~Sakstein and B.~Jain,
Phys. Rev. Lett. \textbf{119} (2017) no.25, 251303
doi:10.1103/PhysRevLett.119.251303 [arXiv:1710.05893
[astro-ph.CO]].


\bibitem{Boran:2017rdn}
S.~Boran, S.~Desai, E.~O.~Kahya and R.~P.~Woodard,
Phys. Rev. D \textbf{97} (2018) no.4, 041501
doi:10.1103/PhysRevD.97.041501 [arXiv:1710.06168 [astro-ph.HE]].


\bibitem{Oikonomou:2021kql}
V.~K.~Oikonomou,
Class. Quant. Grav. \textbf{38} (2021) no.19, 195025
doi:10.1088/1361-6382/ac2168 [arXiv:2108.10460 [gr-qc]].



\bibitem{Oikonomou:2022xoq}
V.~K.~Oikonomou,
Astropart. Phys. \textbf{141} (2022), 102718
doi:10.1016/j.astropartphys.2022.102718 [arXiv:2204.06304
[gr-qc]].


\bibitem{Odintsov:2020sqy}
S.~D.~Odintsov, V.~K.~Oikonomou and F.~P.~Fronimos,
Nucl. Phys. B \textbf{958} (2020), 115135
doi:10.1016/j.nuclphysb.2020.115135 [arXiv:2003.13724 [gr-qc]].




\bibitem{Oikonomou:2020oex}
V.~K.~Oikonomou,
Phys. Rev. D \textbf{103} (2021) no.12, 124028
doi:10.1103/PhysRevD.103.124028 [arXiv:2012.01312 [gr-qc]].






\bibitem{Planck:2018jri}
Y.~Akrami \textit{et al.} [Planck],
Astron. Astrophys. \textbf{641} (2020), A10
doi:10.1051/0004-6361/201833887 [arXiv:1807.06211 [astro-ph.CO]].


\bibitem{BICEP:2021xfz}
P.~A.~R.~Ade \textit{et al.} [BICEP and Keck],
Phys. Rev. Lett. \textbf{127} (2021) no.15, 151301
doi:10.1103/PhysRevLett.127.151301 [arXiv:2110.00483
[astro-ph.CO]].





\bibitem{Benetti:2021uea}
M.~Benetti, L.~L.~Graef and S.~Vagnozzi,
Phys. Rev. D \textbf{105} (2022) no.4, 043520
doi:10.1103/PhysRevD.105.043520 [arXiv:2111.04758 [astro-ph.CO]].

\bibitem{Lyth:1996im}
D.~H.~Lyth,
Phys. Rev. Lett. \textbf{78} (1997), 1861-1863
doi:10.1103/PhysRevLett.78.1861 [arXiv:hep-ph/9606387 [hep-ph]].


\bibitem{LiteBIRD:2022cnt}
E.~Allys \textit{et al.} [LiteBIRD],
PTEP \textbf{2023} (2023) no.4, 042F01 doi:10.1093/ptep/ptac150
[arXiv:2202.02773 [astro-ph.IM]].








\bibitem{Caputo:2024oqc}
A.~Caputo and G.~Raffelt,
PoS \textbf{COSMICWISPers} (2024), 041 doi:10.22323/1.454.0041
[arXiv:2401.13728 [hep-ph]].


\bibitem{Kuster:2008zz}
M.~Kuster, G.~Raffelt and B.~Beltran,
Lect. Notes Phys. \textbf{741} (2008), pp.1-258
doi:10.1007/978-3-540-73518-2

\bibitem{Marsh:2015xka}
D.~J.~E.~Marsh,
Phys. Rept. \textbf{643} (2016), 1-79
doi:10.1016/j.physrep.2016.06.005 [arXiv:1510.07633
[astro-ph.CO]].



\bibitem{Oikonomou:2024aww}
V.~K.~Oikonomou, P.~Tsyba and O.~Razina,
EPL \textbf{145} (2024) no.4, 49001 doi:10.1209/0295-5075/ad239c
[arXiv:2402.02049 [gr-qc]].





\bibitem{Kamionkowski:2015yta}
M.~Kamionkowski and E.~D.~Kovetz,
Ann. Rev. Astron. Astrophys. \textbf{54} (2016), 227-269
doi:10.1146/annurev-astro-081915-023433 [arXiv:1510.06042
[astro-ph.CO]].




\bibitem{Turner:1993vb}
M.~S.~Turner, M.~J.~White and J.~E.~Lidsey,
Phys. Rev. D \textbf{48} (1993), 4613-4622
doi:10.1103/PhysRevD.48.4613 [arXiv:astro-ph/9306029 [astro-ph]].

\bibitem{Boyle:2005se}
L.~A.~Boyle and P.~J.~Steinhardt,
Phys. Rev. D \textbf{77} (2008), 063504
doi:10.1103/PhysRevD.77.063504 [arXiv:astro-ph/0512014
[astro-ph]].



\bibitem{Zhang:2005nw}
Y.~Zhang, Y.~Yuan, W.~Zhao and Y.~T.~Chen,
Class. Quant. Grav. \textbf{22} (2005), 1383-1394
doi:10.1088/0264-9381/22/7/011 [arXiv:astro-ph/0501329
[astro-ph]].



\bibitem{Caprini:2018mtu}
C.~Caprini and D.~G.~Figueroa,
Class. Quant. Grav. \textbf{35} (2018) no.16, 163001
doi:10.1088/1361-6382/aac608 [arXiv:1801.04268 [astro-ph.CO]].




\bibitem{Clarke:2020bil}
T.~J.~Clarke, E.~J.~Copeland and A.~Moss,
JCAP \textbf{10} (2020), 002 doi:10.1088/1475-7516/2020/10/002
[arXiv:2004.11396 [astro-ph.CO]].



\bibitem{Smith:2005mm}
T.~L.~Smith, M.~Kamionkowski and A.~Cooray,
Phys. Rev. D \textbf{73} (2006), 023504
doi:10.1103/PhysRevD.73.023504 [arXiv:astro-ph/0506422
[astro-ph]].




\bibitem{Giovannini:2008tm}
M.~Giovannini,
Class. Quant. Grav. \textbf{26} (2009), 045004
doi:10.1088/0264-9381/26/4/045004 [arXiv:0807.4317 [astro-ph]].

\bibitem{Liu:2015psa}
X.~J.~Liu, W.~Zhao, Y.~Zhang and Z.~H.~Zhu,
Phys. Rev. D \textbf{93} (2016) no.2, 024031
doi:10.1103/PhysRevD.93.024031 [arXiv:1509.03524 [astro-ph.CO]].


\bibitem{Giovannini:2023itq}
M.~Giovannini,
[arXiv:2303.11928 [gr-qc]].






\bibitem{Giovannini:2022eue}
M.~Giovannini,
Eur. Phys. J. C \textbf{82} (2022) no.9, 828
doi:10.1140/epjc/s10052-022-10800-4 [arXiv:2206.08217 [gr-qc]].


\bibitem{Giovannini:2022vha}
M.~Giovannini,
Phys. Rev. D \textbf{105} (2022) no.10, 103524
doi:10.1103/PhysRevD.105.103524 [arXiv:2203.13586 [gr-qc]].


\bibitem{Giovannini:2020wrx}
M.~Giovannini,
Phys. Lett. B \textbf{810} (2020), 135801
doi:10.1016/j.physletb.2020.135801 [arXiv:2006.02760 [gr-qc]].


\bibitem{Giovannini:2019oii}
M.~Giovannini,
Prog. Part. Nucl. Phys. \textbf{112} (2020), 103774
doi:10.1016/j.ppnp.2020.103774 [arXiv:1912.07065 [astro-ph.CO]].


\bibitem{Giovannini:2019ioo}
M.~Giovannini,
Phys. Rev. D \textbf{100} (2019) no.8, 083531
doi:10.1103/PhysRevD.100.083531 [arXiv:1908.09679 [hep-th]].



\bibitem{Giovannini:2014vya}
M.~Giovannini,
Phys. Rev. D \textbf{91} (2015) no.2, 023521
doi:10.1103/PhysRevD.91.023521 [arXiv:1410.5307 [hep-th]].


\bibitem{Giovannini:2009kg}
M.~Giovannini,
PMC Phys. A \textbf{4} (2010), 1 doi:10.1186/1754-0410-4-1
[arXiv:0901.3026 [astro-ph.CO]].



\bibitem{Kamionkowski:1993fg}
M.~Kamionkowski, A.~Kosowsky and M.~S.~Turner,
Phys. Rev. D \textbf{49} (1994), 2837-2851
doi:10.1103/PhysRevD.49.2837 [arXiv:astro-ph/9310044 [astro-ph]].

\bibitem{Giare:2020vss}
W.~Giar\`e and F.~Renzi,
Phys. Rev. D \textbf{102} (2020) no.8, 083530
doi:10.1103/PhysRevD.102.083530 [arXiv:2007.04256 [astro-ph.CO]].


\bibitem{Zhao:2006mm}
W.~Zhao and Y.~Zhang,
Phys. Rev. D \textbf{74} (2006), 043503
doi:10.1103/PhysRevD.74.043503 [arXiv:astro-ph/0604458
[astro-ph]].






\bibitem{Lasky:2015lej}
P.~D.~Lasky, C.~M.~F.~Mingarelli, T.~L.~Smith, J.~T.~Giblin,
D.~J.~Reardon, R.~Caldwell, M.~Bailes, N.~D.~R.~Bhat,
S.~Burke-Spolaor and W.~Coles, \textit{et al.}
Phys. Rev. X \textbf{6} (2016) no.1, 011035
doi:10.1103/PhysRevX.6.011035 [arXiv:1511.05994 [astro-ph.CO]].





\bibitem{Cai:2021uup}
R.~G.~Cai, C.~Fu and W.~W.~Yu,
[arXiv:2112.04794 [astro-ph.CO]].


\bibitem{Odintsov:2021kup}
S.~D.~Odintsov, V.~K.~Oikonomou and F.~P.~Fronimos,
Phys. Dark Univ. \textbf{35} (2022), 100950
doi:10.1016/j.dark.2022.100950 [arXiv:2108.11231 [gr-qc]].






\bibitem{Lin:2021vwc}
J.~Lin, S.~Gao, Y.~Gong, Y.~Lu, Z.~Wang and F.~Zhang,
[arXiv:2111.01362 [gr-qc]].

\bibitem{Zhang:2021vak}
F.~Zhang, J.~Lin and Y.~Lu,
Phys. Rev. D \textbf{104} (2021) no.6, 063515 [erratum: Phys. Rev.
D \textbf{104} (2021) no.12, 129902]
doi:10.1103/PhysRevD.104.063515 [arXiv:2106.10792 [gr-qc]].

\bibitem{Visinelli:2017bny}
L.~Visinelli, N.~Bolis and S.~Vagnozzi,
Phys. Rev. D \textbf{97} (2018) no.6, 064039
doi:10.1103/PhysRevD.97.064039 [arXiv:1711.06628 [gr-qc]].




\bibitem{Pritchard:2004qp}
J.~R.~Pritchard and M.~Kamionkowski,
Annals Phys. \textbf{318} (2005), 2-36
doi:10.1016/j.aop.2005.03.005 [arXiv:astro-ph/0412581 [astro-ph]].

\bibitem{Khoze:2022nyt}
V.~V.~Khoze and D.~L.~Milne,
[arXiv:2212.04784 [hep-ph]].


\bibitem{Casalino:2018tcd}
A.~Casalino, M.~Rinaldi, L.~Sebastiani and S.~Vagnozzi,
Phys. Dark Univ. \textbf{22} (2018), 108
doi:10.1016/j.dark.2018.10.001 [arXiv:1803.02620 [gr-qc]].




\bibitem{Casalino:2018wnc}
A.~Casalino, M.~Rinaldi, L.~Sebastiani and S.~Vagnozzi,
Class. Quant. Grav. \textbf{36} (2019) no.1, 017001
doi:10.1088/1361-6382/aaf1fd [arXiv:1811.06830 [gr-qc]].




\bibitem{ElBourakadi:2022anr}
K.~El Bourakadi, B.~Asfour, Z.~Sakhi, Z.~M.~Bennai and T.~Ouali,
Eur. Phys. J. C \textbf{82} (2022) no.9, 792
doi:10.1140/epjc/s10052-022-10762-7 [arXiv:2209.08585 [gr-qc]].



\bibitem{Sturani:2021ucg}
R.~Sturani,
Symmetry \textbf{13} (2021) no.12, 2384 doi:10.3390/sym13122384




\bibitem{Vagnozzi:2022qmc}
S.~Vagnozzi and A.~Loeb,
Astrophys. J. Lett. \textbf{939} (2022) no.2, L22
doi:10.3847/2041-8213/ac9b0e [arXiv:2208.14088 [astro-ph.CO]].


\bibitem{Arapoglu:2022vbf}
A.~S.~Arapo\u{g}lu and A.~E.~Y\"ukselci,
[arXiv:2210.16699 [gr-qc]].


\bibitem{Giare:2022wxq}
W.~Giar\`e, M.~Forconi, E.~Di Valentino and A.~Melchiorri,
[arXiv:2210.14159 [astro-ph.CO]].



\bibitem{Gerbino:2016sgw}
M.~Gerbino, K.~Freese, S.~Vagnozzi, M.~Lattanzi, O.~Mena,
E.~Giusarma and S.~Ho,
Phys. Rev. D \textbf{95} (2017) no.4, 043512
doi:10.1103/PhysRevD.95.043512 [arXiv:1610.08830 [astro-ph.CO]].

\bibitem{Breitbach:2018ddu}
M.~Breitbach, J.~Kopp, E.~Madge, T.~Opferkuch and P.~Schwaller,
JCAP \textbf{07} (2019), 007 doi:10.1088/1475-7516/2019/07/007
[arXiv:1811.11175 [hep-ph]].




\bibitem{Pi:2019ihn}
S.~Pi, M.~Sasaki and Y.~l.~Zhang,
JCAP \textbf{06} (2019), 049 doi:10.1088/1475-7516/2019/06/049
[arXiv:1904.06304 [gr-qc]].






\bibitem{Khlopov:2023mpo}
M.~Khlopov and S.~R.~Chowdhury,
Symmetry \textbf{15} (2023) no.4, 832 doi:10.3390/sym15040832


\bibitem{Odintsov:2022cbm}
S.~D.~Odintsov, V.~K.~Oikonomou and R.~Myrzakulov,
Symmetry \textbf{14} (2022) no.4, 729 doi:10.3390/sym14040729
[arXiv:2204.00876 [gr-qc]].





\bibitem{Vagnozzi:2020gtf}
S.~Vagnozzi,
Mon. Not. Roy. Astron. Soc. \textbf{502} (2021) no.1, L11-L15
doi:10.1093/mnrasl/slaa203 [arXiv:2009.13432 [astro-ph.CO]].



\bibitem{Pi:2024kpw}
S.~Pi, M.~Sasaki, A.~Wang and J.~Wang,
Phys. Rev. D \textbf{110} (2024) no.10, 103529
doi:10.1103/PhysRevD.110.103529 [arXiv:2407.06066 [astro-ph.CO]].


\bibitem{Oikonomou:2024jqv}
V.~K.~Oikonomou, P.~Tsyba and O.~Razina,
Annals Phys. \textbf{462} (2024), 169597
doi:10.1016/j.aop.2024.169597 [arXiv:2401.11273 [gr-qc]].



\end{thebibliography}
\end{document}